\documentclass[a4paper,11pt]{article}
\pdfoutput=1 

\usepackage{jheppub} 

\usepackage[T1]{fontenc} 
\usepackage{breqn}
\usepackage{url}
\usepackage{derivative}
\usepackage{placeins}
\usepackage{amsmath,mathrsfs}
\usepackage{float}
\usepackage{subcaption}
\usepackage{appendix}
\usepackage[numbers,sort&compress]{natbib}
\usepackage{notoccite}

\title{\boldmath }

\usepackage{xcolor}
\definecolor{c1}{rgb}{0., 0.26, 0.9}
\definecolor{c2}{rgb}{0.5, 0.1, 0.3}

\newcounter{qnumber}

\usepackage{soul} 
\usepackage{cleveref}
\title{Gravitational Waves from Resonant Transitions of Tidally Perturbed Gravitational Atoms}

\author[1]{Antonios Kyriazis,}
\author[2]{Fengwei Yang}

\affiliation[1]{Department of Physics, University of Florida, Gainesville, FL 32611,  U.S.A.}
\affiliation[2]{Institute for Fundamental Theory, Department of Physics,
University of Florida,\\Gainesville, FL 32611, U.S.A. }

\emailAdd{akyriazis@ufl.edu, fengwei.yang@ufl.edu}

\abstract{
Light bosons can form a gravitational atom (GA) around a spinning black hole through the superradiance process. Considering the black hole to be part of a binary system, the tidal potential of the companion periodically perturbs the GA such that an ``atomic'' transition occurs between two of its energy eigenstates. The resonant transition is modeled by the Landau-Zener system, where the orbital frequency of the companion determines the relevant transition.    
In this work, we study a novel quasi-monochromatic gravitational wave signal originating directly from the level transition of the GA in a binary system.
We derive the analytical formulae of both the strain waveform and frequency spectrum of the signal. We further investigate the GA-binary systems that can have a large signal-to-noise ratio in the milli-Hz to deci-Hz frequency band. Using the future space-based gravitational wave observatory DECIGO, we find the signal-to-noise ratio is $\mathcal{O}(10-200)$ for the fine-structure constant $\alpha\simeq 0.3$, host black hole mass $M= 150M_\odot$ and boson mass $\mu \simeq 10^{-13} \rm eV$ at a distance within 100 kpc. Given astrophysical uncertainties about the black hole's initial spin, the degeneracy with other monochromatic signals and the small merger rate at those distances, we conclude that the detection of the signal would be challenging.}

\begin{document}

\maketitle

\section{Introduction}
In black hole physics, superradiance is a process that can extract mass and angular momentum from a spinning black hole. It is triggered when a wave that scatters by the black hole satisfies the condition \cite{Brito_2020}: 

\begin{equation} \label{eqn:superradiance condition}
    \omega < m \Omega_{H},
\end{equation}
where $\omega$ is the frequency of the wave, $\Omega_{H}$ is the angular velocity of the black hole and $m$ is the azimuthal number with respect to the rotation axis. In addition, if the bosonic particles that make up the wave have a small mass, such as axions or axion-like particles \cite{Marsh_2016,dark_matter}, they can form hydrogen-like bound states around the black hole, hence the term ``gravitational atom'' (GA) \cite{Detweiler:1980uk,spectra}. Using the measurements of black holes' spins in X-ray binaries, constraints have been placed on these light particles \cite{Arvanitaki_2011,Arvanitaki_2015,_nal_2021,witte2025steppingsuperradianceconstraintsaxions,hoof2024gettingblackholesuperradiance,mehta2021superradianceexclusionslandscapetype,self-interactions,superradiance_string_theory}. \par 
Searches for exotic bosons by gravitational wave (GW) emissions from the GA in an isolated black hole system have been done in two main channels. These are the annihilations of the GA into gravitons to produce GWs with frequency $2 \mu$, where $\mu$ is the boson's mass, and the spontaneous transition between two superradiant states, producing GWs with frequency equal to the energy difference between the states. These types of signals can be searched for in LIGO and LISA \cite{Arvanitaki_2011,Gravitaitonal_wave_searches,Yang_2023,LIGOScientific:2021rnv}. The inclusion of self-interactions of bosons induces mixing between superradiant states that leads to a GW signal in the deci-Hz frequency range and can also be searched for in ground-based interferometers \cite{self-interactions,deci_Hz,Collaviti_2024,DellaMonica:2025zby}. 

\par
The GA presents rich and intriguing phenomena when perturbed via a companion compact object \cite{Baumann_2019}.
The tidal field of the companion induces resonant transitions between the growing and decaying states of the cloud, causing its demise. The question is: are there any observable signals from these transitions? By using conservation of angular momentum, it has been shown that a transition can back-react to the orbit, causing the orbital frequency to either ``float'' or ``sink'', depending on the type of transition. This can leave distinct imprints on the binary's waveform, a smoking gun signature for the presence of a GA \cite{legacy,Baumann_2020,resonant_history,axion_cloud_backreaction,Ionization,sharp_signals,self_interaction_binary,Guo:2024iye,extreme_mass_ratio,Zhang:2018kib}. Another distinct effect of the back-reaction is the increase of the orbit's eccentricity, while the orbit is within the resonance band. This will drastically alter the distribution of black hole masses and eccentricities in black hole binary systems that are expected to be observed by LISA \cite{Bo_kovi__2024}. Off-resonant mixing between growing and decaying states may also prevent superradiance altogether or cause the decay of the GA, if it has grown, while the back-reaction can, also in this case, leave observable imprints on the inspiral's waveform \cite{Tong_2022}.
\par
Whereas most of the literature on this topic has focused on the signatures that a GA imprints on the inspiral GW signal of the binary, in this paper, we point out a {\it novel} GW signal that comes from the tidally perturbed GA itself. The GWs are generated by the time-varying quadrupole moment of the GA due to the interference of two states during level transitions and we classify them as a monochromatic signal.\par      
We study the waveform and spectrum features of this GW signal in a binary system, including duration, strength and peak frequency. We focus on hyperfine and fine transitions, which occur at small orbital frequencies, when the companion is at a much larger distance than the size of the GA and analytical results for the GW waveform and spectrum can be obtained. \par
The frequency of these types of events can be in DECIGO's and LISA's frequency band, from milli-Hz to deci-Hz, and so we scan the parameter space of black hole masses, mass ratios of companions, and boson masses to determine which type of systems produce the most promising signal-to-noise ratio (SNR). For these computations, we assess the validity of the non-relativistic approximation that we employ throughout for the cloud's wavefunction and the superradiant rates and use the relativistically computed quantities in the parameter space where that approximation fails. 
\par
The paper is structured as follows: in section \ref{section:overview}, we review in brief some basic properties of the GA, the Landau-Zener transition of a two-state system, and the effects of the decay rate of the second mode. In section \ref{section:gr waves}, we present the formalism for the computation of the GW strain waveform and frequency spectrum produced during a given transition and compare our signal to that of the inspiral and of the annihilations of the GA. In section \ref{section:application}, we discuss the detectability of the GW signal from a GA-binary system and show the SNR for various model parameters. We conclude in \cref{section:conclusion} with some remarks on the prospects for future investigations in this topic. 

\section{Gravitational Atoms in isolation and in binaries} \label{section:overview}
\subsection{The Gravitational Atom }
In this section, we give a brief overview of the GA, establishing also the conventions we will be using throughout. An important quantity is the ``fine-structure'' constant of a GA, defined as the ratio of the gravitational radius of the black hole to the Compton wavelength of the boson:
\begin{equation}
    \alpha \equiv \frac{r_{g}}{\lambda_{c}} = \mu M,
\end{equation}
where we use the Planck unit throughout the paper, $G=c=\hbar = 1 $, $\mu$ is the boson mass and $M$ is the host black hole mass.  
In the case where the particle is non-relativistic we can approximate $\omega \simeq \mu$. The angular velocity of the Kerr black hole is given by $\Omega_{H} = \frac{\tilde{a}}{2 r_{+}}$, where $\tilde{a} = \frac{a}{M} \leq 1$ is its dimensionless spin and $r_{+} = M + \sqrt{M^{2} - a^{2}}$ is its outer horizon \cite{Carroll:2004st}. The condition of superradiance \cref{eqn:superradiance condition} can be re-expressed as an inequality for $\alpha$:
\begin{equation}
\label{eq:supercondition}
    \alpha<\frac{m}{2} \frac{\tilde{a}}{1 + \sqrt{1 - \tilde{a}^{2}}}.
\end{equation}
Thus, \cref{eq:supercondition} determines the upper bound of the fine-structure constant $\alpha$ given the spin of the black hole $\tilde{a}$ and the azimuthal quantum number $m$ of the GA.
For maximally spinning black holes, $\tilde{a}=1$ and the $m=1$ states can only grow for $\alpha<0.5$. The saturated spin of the black hole can be obtained by rearranging \cref{eq:supercondition},
\begin{equation}
    \tilde{a}_{\textrm{crit}} = \frac{4 m \alpha}{m^{2} + 4 \alpha^{2}}.
\label{eqn:saturated}\end{equation}

The equation of motion of the non-relativistic scalar field produced by superradiance is described by a Schr\"{o}dinger-like equation in the limit $r \gg M$ and $\alpha \ll 1$ \cite{Detweiler:1980uk}:
\begin{equation}
\label{eq:Schrodinger_eq}
    i \partial_{t} \psi(t,\vec{x}) =  \left( - \frac{\nabla^{2}}{2 \mu} - \frac{\alpha}{r} \right) \psi(t,\vec{x}).
\end{equation}
The size of the cloud is characterized by its Bohr radius $r_{c} = \frac{M}{\alpha^{2}}$. The bound states are given by: 
\begin{equation}
    \psi_{nlm} (t,\vec{x}) = e^{-i (\omega_{nlm} - \mu) t} R_{nl}(r) Y_{lm} (\theta,\varphi),
\label{eqn:ansatz}\end{equation}
where $n,l,m$ are the principal, angular, and azimuthal quantum numbers respectively, $\omega_{nlm}$ is the eigenfrequency of the eigenstate $|nlm\rangle$, which is generally complex due to the purely incoming boundary conditions at the black hole's outer horizon, $R_{nl}(r)$ is the hydrogenic radial wavefunction, and $Y_{lm}(\theta,\varphi)$ are the spherical harmonics.  
This non-relativistic approach holds very well if $\alpha \lesssim 0.1$, while it starts to break down for larger values of $\alpha$ and numerical approaches are required to solve the equation of motion \cite{spectra}.\footnote{By numerically solving the full relativistic equation of motion, one can find that the relativistic wavefunctions and eigenenergies (the real part of the eigenfrequencies) align with the non-relativistic ones if $\alpha\lesssim0.3$ while the decay rate (the imaginary part of the eigenfrequencies) starts to deviate from the non-relativistic one when $\alpha\gtrsim0.1$ (see also \cref{sec:relativistic}).}
\begin{figure}
    \centering
    \includegraphics[width=0.8\linewidth]{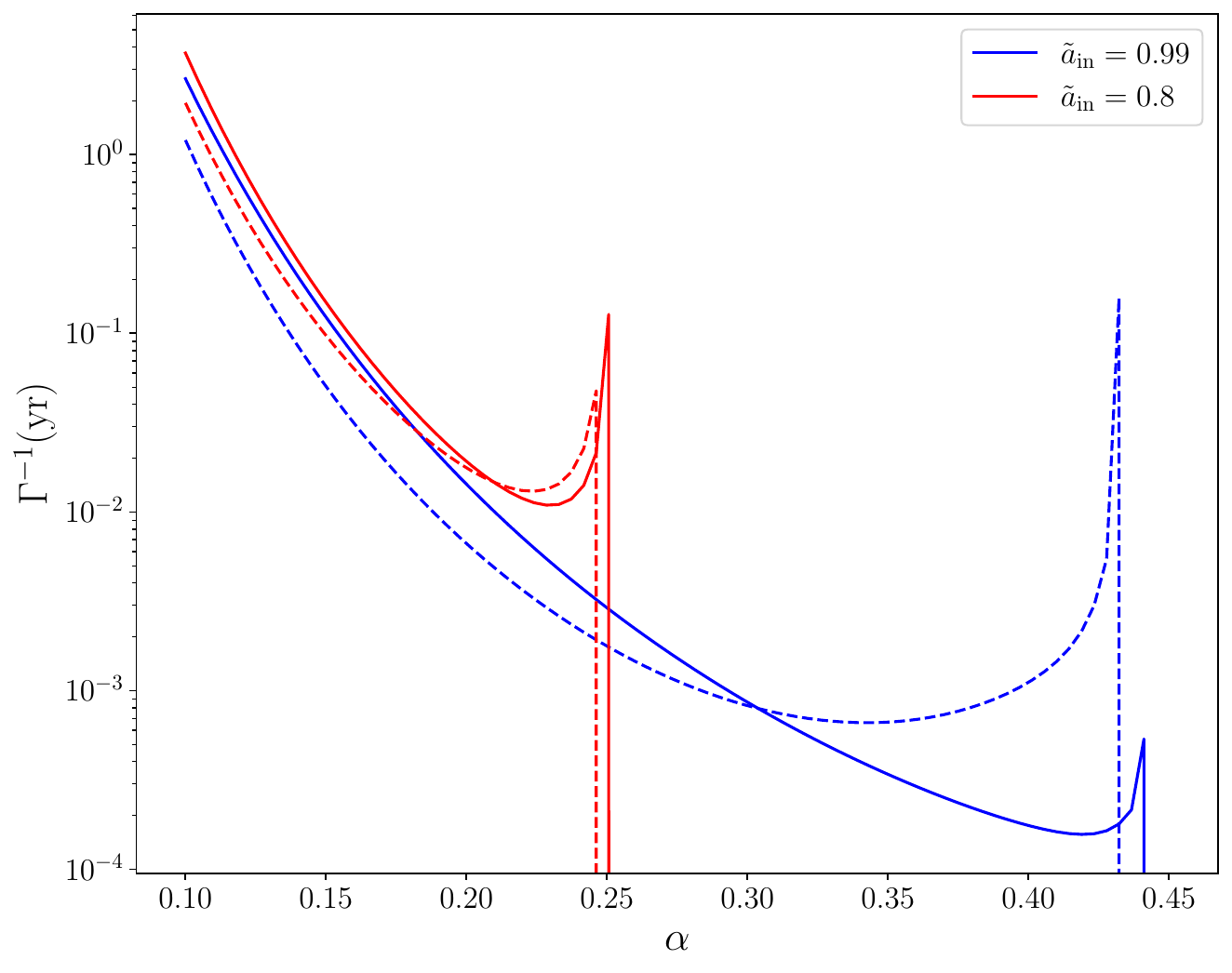}
    \caption{Superradiant timescale in years for the $|211\rangle$ state for two different initial black hole spins as a function of the $\alpha$ parameter. The dashed lines are the non-relativistic approximations \cref{eqn:sr rates}.  The black hole mass is set to $150 M_{\odot}$.}
    \label{fig:Superrad rate}
\end{figure}
The real part of the eigenfrequencies gives the energy levels of the gravitational atom, while the imaginary part gives the instability rates \cite{Detweiler:1980uk,Arvanitaki_2011,spectra}:
\begin{align}
    \label{eqn:energy level}& \omega_{R,nlm} =  \mu \left( 1 - \frac{\alpha^{2}}{2n^{2}} - \frac{\alpha^{4}}{8n^{4}} + \left(\frac{2}{n} - \frac{6}{2l+1} \right) \frac{\alpha^{4}}{n^{3}} + \frac{16\tilde{a} m \alpha^{5}}{n^{3} 2l (2l+1)(2l+2)}  +\mathcal{O}(\alpha^6)\right), \\ 
    \label{eqn:sr rates}& 
    \Gamma_{nlm}\equiv\omega_{I,nlm} = \frac{2 r_{+}}{M} C_{nlm} (\alpha,\tilde{a}) (m \Omega_{H} - \mu) \alpha^{4l +5},
\end{align}
where $C_{nlm}$ are coefficients that can be readily found in \cite{spectra}.
The first two terms in \cref{eqn:energy level} are the same as those of the regular hydrogen atom. The fourth term is what will be relevant for the so-called fine transitions $(\Delta n =0, \Delta l \neq 0)$ while the fifth term, in which the black hole's spin breaks the degeneracy of the third quantum numbers $m$, is relevant for the hyperfine transitions $(\Delta n =0, \Delta l = 0, \Delta m \neq 0)$.
Since we are in the limit $\alpha \ll 1$ and $\Gamma$ strongly depends on $l$, \cref{eqn:sr rates} implies that the fastest growing superradiant levels are those with $l = m $, with the $l=m=1$ being the fastest. An estimate for the growth timescale of the fastest growing mode, around a maximally spinning black hole, is
\begin{equation}
    \Gamma^{-1}_{211} \simeq 0.3 \,\text{days} \left( \frac{M}{150 M_{\odot}} \right) \left( \frac{0.3}{\alpha}\right)^{9} \left( \frac{0.6}{1-2 \alpha}\right)^{3}.
\label{eqn:supperad growth timescale}\end{equation}

\Cref{fig:Superrad rate} shows the superradiant timescale for the $|211\rangle$ state for two different initial black hole spins, $\tilde{a}_{\rm in}=0.8,\,0.99$, comparing the relativistic result using the numerical code in \cite{hoof2024gettingblackholesuperradiance,hoof_git}, with the non-relativistic result in \cref{eqn:sr rates}. The results agree up to $\mathcal{O}(1)$ factors for $\alpha\lesssim 0.25$ with $\tilde{a}_{\rm in}=0.8$ and for $\alpha\lesssim 0.30$ with $\tilde{a}_{\rm in}=0.99$. In general, they diverge for large values of $\alpha$ and large spins. We also note that the smaller the black hole spin is, the smaller the range of $\alpha$ for which the superradiant condition is satisfied. Based on this result, we restrict the analysis to $\alpha<0.42$ in what follows.

\subsection{The perturbed Gravitational Atom}
In this section, we start to consider the GA in a binary system setup and summarize the results from the literature that are relevant to the later analysis. 
The companion in the binary system induces a tidal field $V_{\ast}(t,\vec{r})$ on the GA that will cause it to undergo transitions between states. We will assume that the companion is orbiting at a large distance from the host black hole so that we can treat the tidal field perturbatively. The exact form of the mixing between the states can be found in \cite{Baumann_2019,Baumann_2020,axion_cloud_backreaction} and is summarized in \cref{appendix: tidal} for reference. For our purposes, it is important to note that if the companion is far from the GA, the dominant multipole moment of the tidal field is the {\it quadrupole} moment, $l_*=2$. This leads to mixing between two states and the following selection rules are derived: 
\begin{align}
\label{eq:selection_rule}
    & m_{f} - m_{i} = m_{\ast}, \\
    & |l_{i} - l_{f}| \leq 2 \leq l_{i} + l_{f}, \\
    & l_{i} + l_{f} = 2p -2, p \in \mathbb{Z},
\end{align}
where $i,f$ represent the initial and final states respectively, and $m_*$ is the azimuthal number of the spherical harmonics of the tidal field quadrupole.
Throughout the paper, we make the following two assumptions to simplify the analysis and facilitate the analytical implementations: 
\begin{itemize}
    \item The orbit is quasi-circular. 
    \item The companion is on the same equatorial plane as the cloud.
\end{itemize}
With these assumptions, only states that satisfy $m_{\ast}= \pm 2 = m_{f} - m_{i}$ couple to each other as demanded by the selection rule, and the mixing between two different states takes a simple form\,\footnote{Relaxing the assumption regarding the orbit being on the GA's equatorial plane introduces higher harmonics in \cref{eqn:matrix element}.}:
\begin{equation}
\langle \psi_{f} |V_{\ast} (t,\vec{r}) | \psi_{i} \rangle  = \eta e^{-i m_{\ast} \varphi_{\ast}},
\label{eqn:matrix element}\end{equation}
where $\eta$ denotes the amplitude of the mixing with its explicit expression given in \cref{eqn:eta} and $\varphi_{\ast}$ is the phase of the binary, whose derivative is the frequency of the orbit, $\dot{\varphi_*}(t) = \pm \Omega(t)$
\cite{Baumann_2019,Baumann_2020}. The plus (minus) sign is for co(counter)-rotating orbits. The matrix element in \cref{eqn:matrix element} is akin to a periodic, external driving force, and a resonance is expected to take place when the frequency of this force is equal to the energy splitting of the two states $|m_*|\Omega = \Delta \omega_{R}\equiv\omega_{R,i}-\omega_{R,f}$. \par

The orbit of the companion evolves very slowly during the transition\footnote{ Back-reaction effects affect the transition in two ways: 1. they shift the resonance frequency of the LZ transition \cite{axion_annihilation}; 2. they vary the change rate of the orbital frequency. For the $|211\rangle$ hyperfine transition we considered throughout the paper, the first effect renders the transition non-adiabatic for equal mass binaries, while we will show in \cref{section:application} that our signal is mostly relevant for intermediate mass ratios. The second effect stalls the orbital frequency around the hyperfine frequency, increasing the duration of the resonance \cite{Baumann_2020}. When the decay rate is included, the resonance may break if the occupation number of the first state drops below a threshold \cite{resonant_history}. Even though back-reaction should be included systematically, we have checked that the resonant breaking effect is small for the $|211\rangle \rightarrow |21 \textrm{-}1\rangle$ transition, in the parameter space of interest.} due to the emission of GWs, so that the frequency can be linearized as a function of time $t$ \cite{Baumann_2020}:
\begin{equation}
    \Omega(t) =\Omega_{0} + \gamma t,
\label{eqn:linear freq}\end{equation}
where $\Omega_{0}$ is the reference frequency at $t=0$ and $\gamma$ is the change rate of the orbital frequency due to GW emission,
\begin{equation}
    \frac{\gamma}{\Omega^{2}_{0}} = \frac{96}{5} \frac{q}{(q+1)^{1/3}} (M \Omega_{0})^{5/3},
\label{eqn: gamma}\end{equation}
where $q$ is the mass ratio of the companion.
The linear approximation of the orbital frequency is valid for the time within $-\Omega_0/\gamma \lesssim t \lesssim \Omega_0/\gamma$.
The Hamiltonian that governs the evolution of the coefficients of the initial and final states of a two-state system, $c_{i}(t)$ and $c_{f} (t)$, that are mixed via the perturbing potential $V_*$, is given by
\begin{equation}
\mathcal{H} = 
    \begin{pmatrix}
 -\Delta \omega_{R}/2   &  \eta e^{i \Delta m \varphi_*(t)}  \\
    \eta e^{- i \Delta m \varphi_*(t)}  & \Delta \omega_{R}/2 - i |\Gamma|
\end{pmatrix},
\end{equation}
where $\Delta \omega_R = \omega_f-\omega_i$, 
$\Delta m = m_f-m_i$, and $\Gamma$ is the instability rate of the final state which characterizes the final-state decay due to black hole absorption (see \cref{eqn:sr rates}).
By performing a unitary transformation of $\mathcal{H}$ to the co-rotating (or dressed) frame, one can eliminate the fast oscillations of the off-diagonal elements, and the coefficients of two states transform accordingly
\begin{equation}
\begin{pmatrix}
    c_{i}(t) \\
    c_{f}(t)
\end{pmatrix}
= 
    \begin{pmatrix}
          e^{i \Delta m \varphi_*(t)/2} &  0  \\
          0  & e^{- i \Delta m \varphi_*(t)/2} 
    \end{pmatrix}
    \begin{pmatrix}
    d_{i}(t) \\
    d_{f}(t)
\end{pmatrix}.
\label{eqn:cs}\end{equation}
Since it has been used widely in the literature, we provide the Hamiltonian in this dressed frame directly in \cref{appendix:Hamiltonian}. When setting the reference orbital frequency to match the energy splitting of the transition $\Omega_{0} = {\pm} \frac{\Delta \omega_{R}}{\Delta m}$ and applying the linear orbital frequency approximation, in \cref{eqn:linear freq}, the Hamiltonian $\mathcal{H}$ is reduced to the well known Landau-Zener system \cite{Landau,Zener}. The main point is that a transition will take place at frequency $\Omega_{0}$. In this paper, we will primarily deal with hyperfine and fine transitions, whose orbital frequency can be found from \cref{eqn:energy level},
\begin{align}  
    \label{eqn:Omega hyperfine}\Omega_{0,\textrm{hyp}}  
    &=\frac{64 m_i\alpha^{7}}{Mn^{3} 2l (2l+1)(2l+2)(m_i^2+4\alpha^2)}, 
    \\  
   \label{eqn:Omega fine} \Omega_{0,\textrm{fine}} &= \frac{\Delta l}{\Delta m}\frac{12 \alpha^{5}}{M n^{3} (2l_i+1)(2l_f+1)}.
\end{align}
where we assumed for hyperfine transitions that the black hole's spin is given by \cref{eqn:saturated}, and the subscript of $l$ and $m$ represents the associated initial or final state.
An advantage of working with hyperfine and fine transitions is that they take place when the companion is far away from the GA, and so keeping only the quadrupole term in the tidal field is justified \cite{hyperfine_trans_are_favoured}. We will consider hyperfine transitions from the $|211\rangle$ state and fine transitions from the $|322\rangle$ state. The selection rules that we discussed above demand $|211\rangle\rightarrow |21\text{-}1\rangle$ and $|322\rangle \rightarrow |300\rangle$. Since $\Delta m = -2$ and $\Delta E <0$, only co-rotating orbits can induce these transitions, which we will consider throughout the paper. \par
In addition, the orbital frequency of these transitions happens to fall in the $\rm mHz - \rm Hz$ range, depending on the particular transition, black hole mass and $\alpha$. For example, the frequency for the $|211\rangle\rightarrow |21\text{-}1\rangle$ transition, $f_0\equiv\Omega_0/(2\pi)$:
\begin{equation}
   f_{0,211} \simeq 10 {\rm mHz}\,\left(\frac{\alpha}{0.3} \right)^7 \left(\frac{150 M_\odot}{M}\right) \left(\frac{1.36}{1+4 \alpha^{2}} \right).
\end{equation}
Similarly, for the fine transition $|322\rangle \rightarrow |300\rangle$, 
\begin{equation}
 f_{0,322} \simeq 46  \textrm{mHz} \left(  \frac{\alpha}{0.3}\right)^{5} \left( \frac{150 M_{\odot}}{M}\right).
\end{equation}

\par
An important parameter that characterizes the transition is the {\it adiabaticity} parameter:
\begin{equation}
    z \equiv \frac{\eta^{2}}{|\Delta m|\gamma}.
\label{eqn: adiabaticity}\end{equation}
For the $|211\rangle \rightarrow | 21\textrm{-}1 \rangle$ hyperfine transition, the typical value of $z$ is 
\begin{equation}
   z \simeq 0.7  \left( \frac{1.36}{1+4\alpha^{2}}\right)^{1/3} \left(\frac{q}{1/150}\right) \left( \frac{0.3}{\alpha}\right)^{11/3}.
\label{eqn:z hyperfine}
\end{equation}

\begin{figure*}[t!]
    \centering
    \begin{subfigure}[t]{0.8\textwidth}
        \centering
        \includegraphics[width=0.99\linewidth]{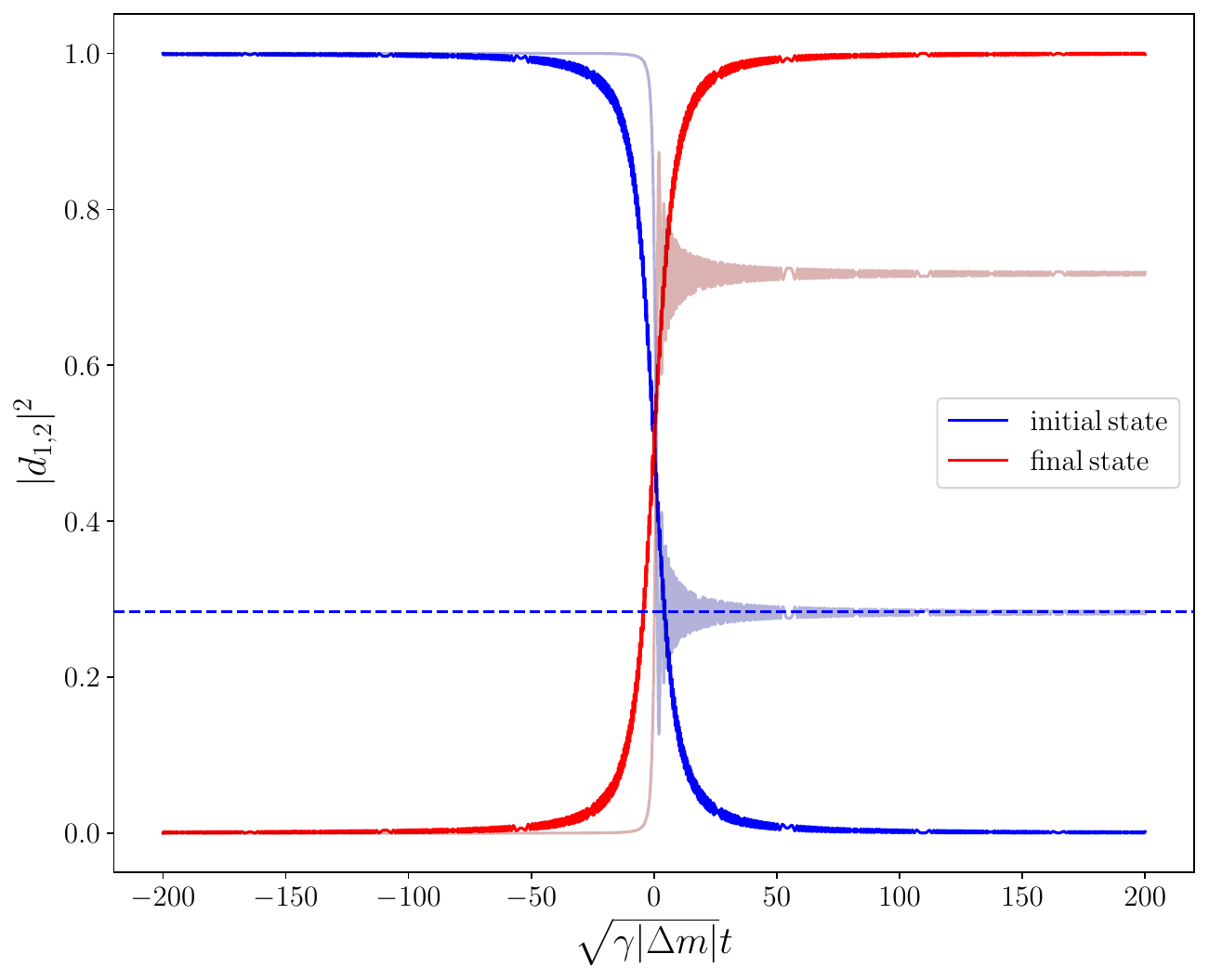}
    \end{subfigure}%
    ~
\caption{The adiabatic and non-adiabatic transitions for $z=20$ (bold) and $z=0.2$ (pale). The horizontal dashed blue line shows the residual population $e^{-2 \pi z}$ of the initial state for the second case.}
   \label{fig:hyperfine}
\end{figure*}

The remaining population of the first state long after the transition is given by $|d_{1} (t \rightarrow \infty)|^{2} = e^{-2 \pi z}$. If the adiabaticity $z \gg 1$, the transition is adiabatic: the two states exchange populations after the transition. When $z \lesssim 1$, the transition is non-adiabatic and only a fraction of the bosons is transferred. The evolution of the states in these two cases is shown in \cref{fig:hyperfine}. \par
\par
In the co-rotating frame, the Hamiltonian is transformed to a simpler form $\mathcal{\bar{H}}$ (see \cref{eqn:Hamiltonian}) and the coefficients $d_1$ and $d_2$ evolve according to
\begin{equation}
\label{eq:eomOfd}
    i\frac{dd_a}{dt}=\sum_b \mathcal{\bar{H}}_{ab}(t)d_b,~~a,b=1,2.
\end{equation}
The solutions with the initial conditions $|d_{1} (t \rightarrow -\infty)|^{2} = 1 $ and $|d_{2} (t \rightarrow -\infty)|^{2} = 0$, ignoring the decay rate of the final state $\Gamma=0$, have an analytical form in terms of the parabolic cylinder function \cite{D},
\begin{align}
\label{eq:coefficient_d1}
    & d_{1}(t) = e^{- \frac{\pi z}{4}} D_{iz} (e^{\frac{3 i \pi}{4}} \sqrt{|\Delta m| \gamma} t), \\
    \label{eq:coefficient_d2}
    & d_{2}(t) = \sqrt{z} e^{-\frac{\pi z}{4}} D_{iz-1} (e^{\frac{3i\pi}{4}} \sqrt{|\Delta m| \gamma} t).
\end{align}

\subsection{The decay rate of the final state}\label{sec: decay rate}
So far, we have ignored the decay of the final state into the black hole. However, if the mixing between the states is subdominant compared to the decay rate $\Gamma$, i.e. $\eta \ll |\Gamma|$, the evolution of the coefficients $d_{1}$ and $d_{2}$ will be significantly different from what we laid out above.  In \cref{eqn:decay211} we evaluate the ratio of the mixing strength $\eta$ with the decay rate $|\Gamma|$ \cite{extreme_mass_ratio}. The latter is calculated assuming that the black hole is at the saturated spin of the corresponding initial state and we may approximate it as $|\Gamma_{21{\text-}1}| \simeq \alpha^{10}/(6 M)$, where we have used the analytical expressions that can be found in \cite{Detweiler:1980uk,Arvanitaki_2011,spectra} in the limit $\alpha \ll 1$. The relevant ratios are
\begin{equation}                         
 \label{eqn:decay211}   \frac{\eta_{211 \rightarrow 21{\text-}1}}{|\Gamma_{21{\text-}1}|} \simeq 6 \times 10^{-3} \left( \frac{q}{1/150} \right) \left( \frac{\alpha}{0.3}\right) \left( \frac{1.36}{1+4\alpha^{2}} \right)^{2} ,
\end{equation} 
and therefore, the decay of the second mode needs to be taken into account for a wide range of parameters and especially when small mass ratios are considered (this ratio becomes even smaller with the general relativistic computation of $|\Gamma|$ \cite{hoof2024gettingblackholesuperradiance,witte2025steppingsuperradianceconstraintsaxions,Siemonsen:2022yyf}). There are analytical results for the coefficients $d_{1}$ and $d_{2}$ in terms of the parabolic cylinder functions that can be readily used \cite{decayrate1,decayrate2}:
\begin{align}
  \label{eqn:d1 general eta Gamma} & d_{1}(t) = e^{-\frac{|\Gamma| t}{2} - \frac{\pi z}{4}} D_{iz} \left(e^{\frac{3i\pi}{4}}\left(\sqrt{\gamma |\Delta m|} t -i \frac{|\Gamma|}{\sqrt{\gamma |\Delta m|}} \right) \right), \\ & \label{eqn: d2 general eta Gamma}
   d_{2}(t) = e^{-\frac{|\Gamma| t}{2} - \frac{\pi z}{4}} \sqrt{z} D_{iz-1} \left(e^{\frac{3i\pi}{4}}\left(\sqrt{\gamma |\Delta m|} t -i \frac{|\Gamma|}{\sqrt{\gamma |\Delta m|}} \right) \right).
\end{align}
Specializing to the case where $\eta \ll \Gamma$, we can simplify these to \cite{extreme_mass_ratio}:
\begin{align}
    \label{eqn: d1^2 gamma>> 1/dt} &
    |d_{1} (t )|^{2} = \text{exp}\left[-z \left( \pi+2\text{tan}^{-1} \left( \frac{|\Delta m| \gamma t}{|\Gamma|}\right)\right) \right],   \\
    \label{eqn: d2 gamma>> 1/dt} & 
    d_{2} ( t) = - \frac{i\eta}{i|\Delta m| \gamma t + |\Gamma|} d_{1}( t). 
\end{align}
\begin{figure}[t!]
    \centering
        \includegraphics[width=0.74\linewidth]{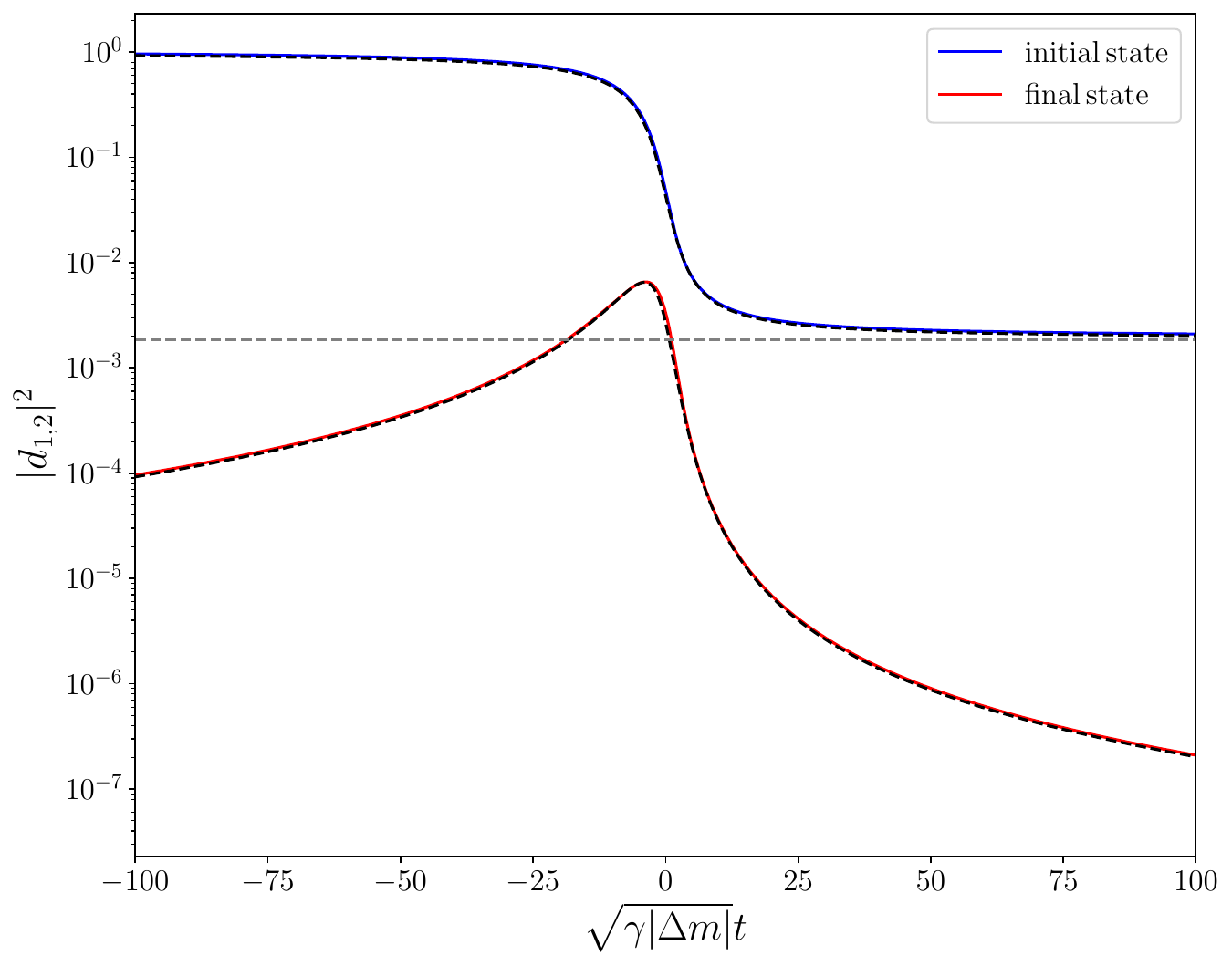}
    \caption{Comparison of the numerical solution (solid) of \cref{eqn:Hamiltonian} and the  analytical results in \cref{eqn: d1^2 gamma>> 1/dt} (dashed). We have chosen $z=1$ and $\frac{\eta}{|\Gamma|} = 0.25$. The horizontal grey dashed line is the asymptotic value $e^{-2 \pi z}$.}
    \label{fig:d1 large Gamma evolution}
\end{figure}

In the limit $|d_{1} \left( t \rightarrow \infty \right)|^{2} \rightarrow e^{-2 \pi z}$, we recover the same result as in the Landau-Zener system for the initial state. In the $t \rightarrow \pm \infty$ limit, the occupation number of the second state is zero, while at $t=0$ it is suppressed by the factor $ \frac{\eta}{|\Gamma|}$, a reflection of the fast decay rate of the bosons. In \cref{fig:d1 large Gamma evolution}, we plot the numerical and the analytical solutions \cref{eqn: d1^2 gamma>> 1/dt,eqn: d2 gamma>> 1/dt} for a certain choice of the ratio $\frac{\eta}{|\Gamma|}$. The agreement is very good between the two. We notice that the oscillations in \cref{fig:hyperfine} that are observed in the $z \lesssim 1$ case are now suppressed. \par

\section{Gravitational waves from transitions }\label{section:gr waves}
In this section, we present the key results of our work, which is the derivation of the GW strain from the tidal perturbation of the GA by the companion object, along with the frequency spectrum. The signal's peak frequency is set by the energy splitting of the level transition, the adiabaticity parameter $ z$ and the decay rate of the second state $|\Gamma|$; the frequency change rate is determined by the orbital frequency change rate of the binary; the amplitude modulation of the strain waveform $h(t)$ is determined by the LZ transition and the evolution of the coefficients $d_{1}$ and $d_{2}$. \par
To start, we employ the quadrupole formula of GW emission, which is derived in detail in \cref{sec:quadrupole}:
\begin{equation}
    h^{\rm TT}_{ij}({\bf x},t)\simeq\frac{2}{r}\Lambda_{ij,lm}(\hat{\bf n})\ddot{M}_{lm},
\label{eqn:quadrupole formula h}\end{equation}
where $M_{ij}$ are the quadrupole moments of the GW source, given by:
\begin{equation}
    M_{i j } = \int d^{3}x \rho(t,\vec{x}) x_{i} x_{j},
\label{eqn:quadrupole}\end{equation}
$\Lambda_{ij,lm}$ is a tensor that depends only on the GW propagation direction $\hat{\bf n}$ and projects the quadrupole moments of the source onto the transverse-traceless (TT) gauge, and $r$ is the distance to the source. We work in the TT gauge and omit the superscript of the strain in the following. \par

The quadrupole formula is valid in the limit where the wavelength of the emitted radiation is much larger than the size of the source, which can be justified once we derive the frequency of the GW signal, and is valid throughout our parameter space. \par

For a real scalar field $\phi$ with mass $\mu$, the energy density is given by
\begin{equation}
    \rho(t,\vec{x}) = \frac{\dot{\phi}^{2}}{2} + \frac{(\nabla \phi)^{2}}{2} + \frac{\mu^{2} \phi^{2}}{2}.
\end{equation}
In the non-relativistic limit $\alpha \ll 1$, we express the field $\phi$ in terms of the wavefunction $\psi$ by separating the fast-oscillation mode $e^{-i\mu t}$ from it,  
\begin{equation}
    \phi(t,\vec{x}) = \frac{1}{\sqrt{2 \mu}} \left(  \psi( t,\vec{x})e^{-i \mu t} + c.c \right),
\end{equation}
for which it holds that $|\partial_{t} \psi| \ll \mu |\psi|$ and $|\nabla \psi| \ll \mu |\psi|$. From \cref{eqn:ansatz}, one can find $|\partial_{t} \psi| \sim \mu \alpha^{2} $ and $|\nabla \psi| \sim \mu \alpha |\psi|$ and therefore the inequalities are satisfied when $\alpha \ll 1$. In these limits, the energy density of $\phi$ is dominated by 
\begin{equation}
    \rho(t,\vec{x}) \simeq \mu |\psi (t,\vec{x})|^{2} + \,...
\end{equation}
where the dots include terms of $\mathcal{O}(\alpha^{2})$ or higher.
\par

The state of the GA $\psi(t,\vec{x})$ is in general a superposition of the two states that participate in the transition
\begin{equation}
     \psi(t,\vec{x}) = \sqrt{N} \left( c_{1}(t) \psi_{1}(\vec{x}) + c_{2}(t) \psi_{2}(\vec{x}) \right),
\end{equation}
where $N$ is the number of axions. The energy density is given by:
\begin{equation}
    \rho(t,\vec{x})=  M_{c} \left( |c_{1}(t)|^{2} |\psi_{1}|^{2} + |c_{2}(t)|^{2}|\psi_{2}|^{2} + 2 \Re\left( c^{\ast}_{1}(t) c_{2}(t) \psi^{\ast}_{1} \psi_{2} \right) \right),
\label{eqn:density}\end{equation}
where $M_{c} = \mu N$ is the mass of the cloud. We stress that the usual normalization condition $|c_{1}(t)|^{2} + |c_{2}(t)|^{2} = 1$ does not apply here because of the decay rate of the second state, in which final-state bosons are dissipated by the black hole absorption.

\par
For the hyperfine $|211\rangle \rightarrow |21\text{-}1\rangle$ transition, 
the relevant quadrupole moments are
\begin{align}
   & M_{11}(t) = 12M_{c} r^{2}_{c} \left(1 - \Re \left[d^{\ast}_{1}(t_{\rm re}) d_{2}(t_{\rm re}) e^{-2i\Delta m \varphi(t_{\rm re})}  \right] \right), \\ &
   M_{22}(t) = 12M_{c} r^{2}_{c} \left(1 + \Re \left[d^{\ast}_{1}(t_{\rm re}) d_{2}(t_{\rm re}) e^{-2i\Delta m \varphi(t_{\rm re})}  \right] \right), \\ &
   M_{12}(t) = -12 M_{c} r^{2}_{c} \Im \left[d^{\ast}_{1}(t_{\rm re}) d_{2}(t_{\rm re}) e^{-2 i \Delta m \varphi(t_{\rm re})} \right],  \\ &
   M_{33}(t) = 6 M_{c} r^{2}_{c} \left( |d_{1}(t_{\rm re})|^{2} + |d_{2}(t_{\rm re})|^{2} \right) , 
\end{align}
where $\Re[\,]$ denotes the real part while $\Im[\,]$ denotes the imaginary part, we used the wavefunctions of the hydrogen atom \cite{Shankar:102017}, the unitary matrix from \cref{eqn:cs} to express the coefficients in the co-rotating frame, and defined the retarded time $t_{\rm re} \equiv t - r$, with $r$ the distance between the GA and the observer. The $M_{13}$ and $M_{23}$ coefficients are zero due to the integral over the azimuthal angle.  \par
The second time derivatives of the above quadrupole moments will determine the GW strain from the transition, which eventually are dependent on the second time derivatives of $d_1$ and $d_2$. The equations of motion for $d_1$ and $d_2$ in \cref{eq:eomOfd} can be used to re-express their second time derivatives in terms of the known solutions $d_1$ and $d_2$. The leading order term of $\ddot{M}_{ij},\,i,j=1,2,3$, can also be obtained in the following way: in the $\eta \ll |\Gamma|$ limit, when a time derivative acts on $d_{1,2}$ given by \cref{eqn: d1^2 gamma>> 1/dt,eqn: d2 gamma>> 1/dt}, the result is of order $ \mathcal{O} \left( \frac{\gamma}{|\Gamma|} d_{1,2}  \right)$. When the time derivative acts on the exponential function $e^{-2i\Delta m\varphi(t_{\rm re})}$, it brings down a factor of $\Omega \sim \Omega_{0}$. By factoring out $\Omega^{2}_{0}$, the remaining terms are of order $\mathcal{O} \left( \frac{\gamma^{2}}{\Omega^{2}_{0} |\Gamma|^{2}}, \frac{\gamma}{\Omega_{0} |\Gamma|} \right)$, which scale like $\mathcal{O} \left( \alpha^{52/3}, \alpha^{26/3} \right)$ for the $|211\rangle \rightarrow |21\text{-}1\rangle$ transition, so they can be ignored. In the leading order, the relevant second-order derivatives are
\begin{align}
      & \ddot{ M}_{11}(t)   = - \ddot{ M}_{22}(t) = 12 M_{c} r^{2}_{c} \Omega_0^{2} \Re \left[ e^{-2 i \Delta m \varphi(t_{\rm re})} Q(t_{\rm re}) \right], \\ 
    & \ddot{ M}_{12} (t)  = 12 M_{c} r^{2}_{c} \Omega_0^{2} \Im \left[ e^{-2 i \Delta m \varphi(t_{\rm re})} Q(t_{\rm re}) \right], \\ & \label{eqn:Q}
    Q(t) = 4 |\Delta m|^{2} d^{\ast}_{1}(t_{\textrm{re}}) d_{2}(t_{\textrm{re}}) +\mathcal{O}(\frac{\gamma}{\Omega_{0} |\Gamma|}),
\end{align}
where we have ignored $\ddot{M}_{33}$ because it is suppressed by $\mathcal{O}\left( \frac{\gamma^{2}}{\Omega^{2}_{0}|\Gamma|^{2} }\right)$ relative to the other components. $Q(t)$ can be found analytically by \cref{eqn: d1^2 gamma>> 1/dt,eqn: d2 gamma>> 1/dt} in the $\eta \ll |\Gamma|$ limit. In this case, it is straightforward to find that the maximum value of $|Q|$ occurs at $t_{\rm max} = - \frac{2 z |\Gamma|}{|\Delta m|\gamma} = - \frac{ z |\Gamma|}{\gamma} $. This result persists for the more general formula as long as $\eta \lesssim |\Gamma|$. 
\begin{figure}
\centering
\begin{subfigure}{.50\textwidth}
  \centering
  \includegraphics[width=0.98\linewidth]{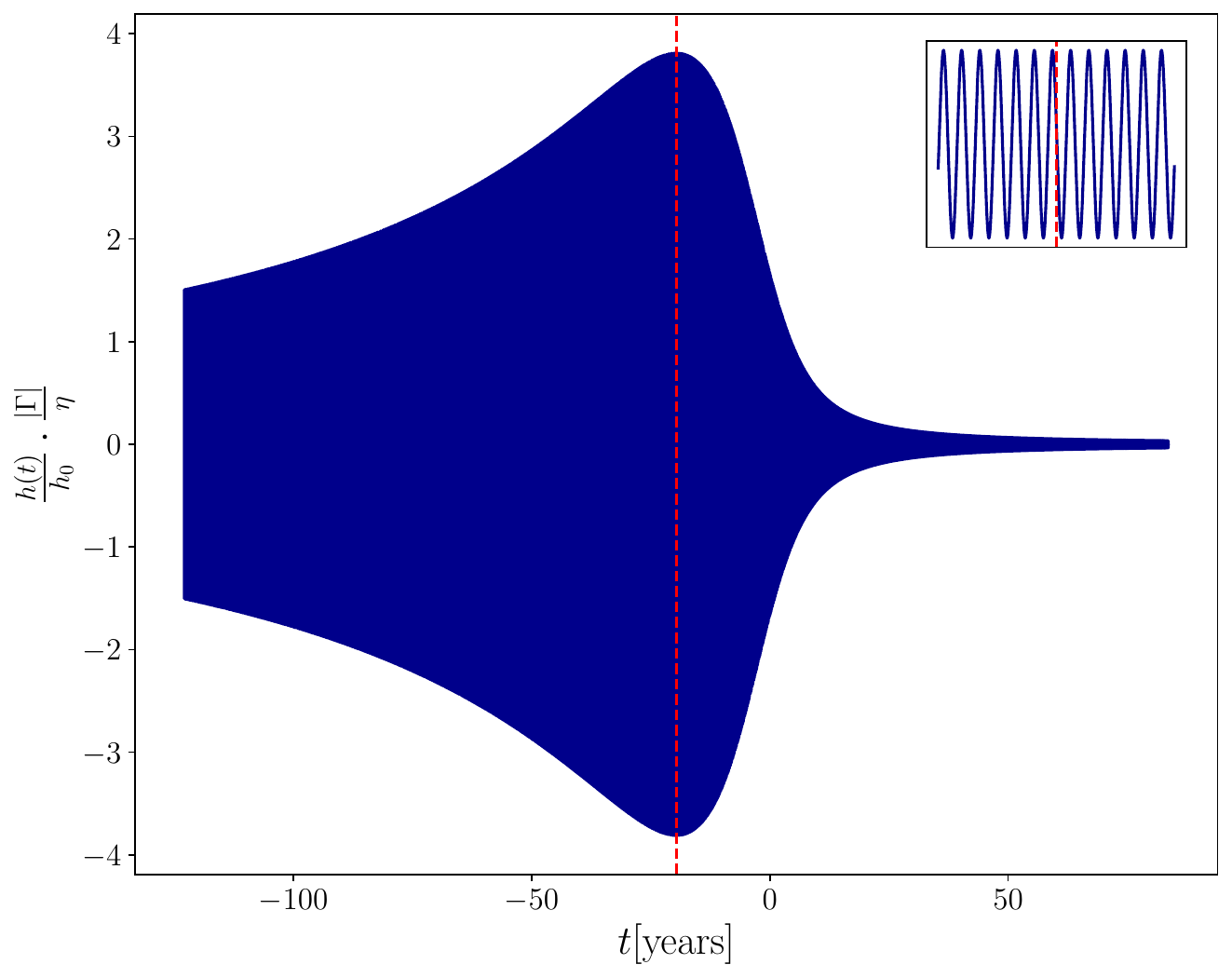}
  \end{subfigure}%
  \hfill
  \begin{subfigure}{.50\textwidth}
  \centering
  \includegraphics[width=1.1\linewidth]{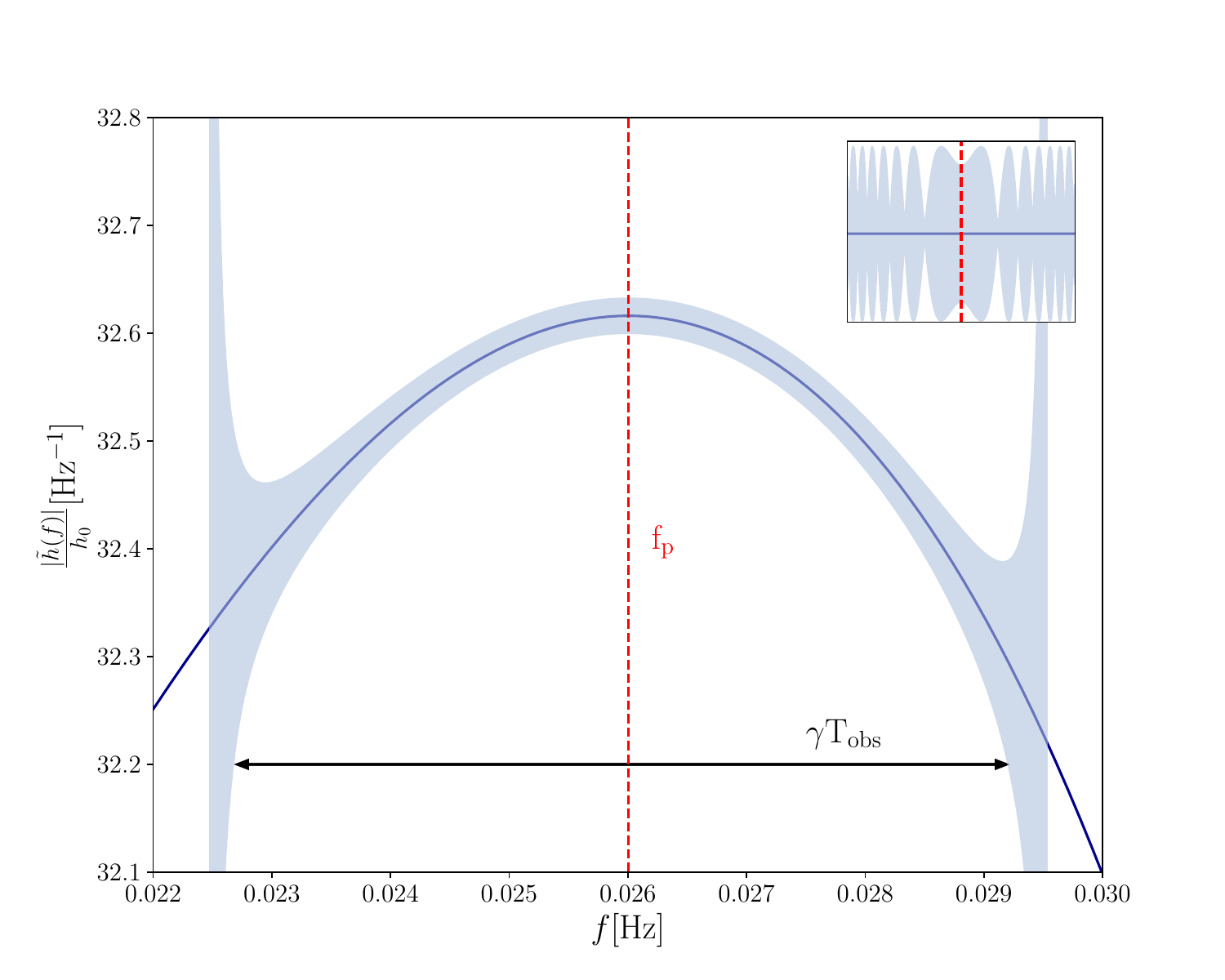}
\end{subfigure}%
 \caption{\label{fig: strain freq time} (left) The plus polarization of the GW signal \cref{eqn: plus strain} versus time for the $|211\rangle \rightarrow |21\textrm{-}1\rangle$ transition. We have chosen $\alpha = 0.3$, $q=1/150$, and $M = 150 M_{\odot}$. The red dashed line is set at $t_{\rm max} = - z |\Gamma|/\gamma$. The inset shows the oscillations around $t_{\rm max}$. (right) The frequency spectrum $|\tilde{h}(f)|/h_{0}$ versus frequency for the same parameters. The solid dark blue line is given by the stationary phase approximation, while the pale blue line is obtained by the short-time Fourier transform. The red dashed vertical line shows the peak frequency of the signal \cref{eqn:peak freq}, while the inset shows the fast oscillations around that frequency. The black arrow shows the frequency width, which is equal to $\gamma T_{\rm obs}$. } 
\end{figure} 

The strain of the plus- and cross-polarized GW is thus given by \cite{Maggiore:2007ulw},
\begin{align}
     \label{eqn: plus strain}h_{+,211}(t) &= h_{0} \frac{1+\cos^{2}(\iota)}{2} \Re \left[ e^{-2 i\Delta m \varphi(t_{\rm re})} Q(t_{\rm re}) \right], 
     \\    h_{\times,211} (t) & = h_{0} \cos(\iota) \Im \left[ e^{-2 i \Delta m \varphi(t_{\rm re})} Q(t_{\rm re}) \right],
\end{align}
where $\iota$ is the angle between the line-of-sight of the observer and normal to the orbital plane of the system. The signal is therefore composed of a fast oscillating exponential term that contains information about the phase of the binary and an amplitude modulation $Q$ that contains the details of the transition.  
The factor $h_{0}$ is given by
\begin{equation}
    h_{0} = \frac{24  M_{c} r^{2}_{c} \Omega^{2}_{0}}{r} = 24 \frac{q_{c} M}{r} \frac{1}{\alpha^{4}} (M \Omega_{0})^{2},
\label{eqn:amp}\end{equation}
where $q_{c}  \equiv \frac{M_{c}}{M}$ is defined as the GA mass ratio. $q_c$ can in general be computed numerically using the SupeRrad package \cite{Siemonsen:2022yyf,May:2024npn}. For example, $q_{c} \simeq 0.085$ for $\alpha=0.3$ and $\tilde{a}_{\rm in} = 0.99$ at saturation. The characteristic value of $h_{0}$ for the benchmark parameters is
\begin{equation}
    h_{0} = 5 \times 10^{-23} \left( \frac{q_{c}}{0.085} \right) \left( \frac{M}{150 M_{\odot}} \right) \left( \frac{100 \textrm{kpc}}{r} \right) \left( \frac{\alpha}{0.3}\right)^{10}.
    \label{eqn:amp-scaling}
\end{equation}

The GW strain waveform is shown in \cref{fig: strain freq time} (left) for the benchmark parameters, $\alpha = 0.3$, $q=1/150$, and $M = 150 M_{\odot}$. The GW signal is quasimonochromatic with its oscillation frequency set by $2|\Delta m|f_0$, where $f_0$ is the orbital frequency of the binary, because of the exponential factor $e^{-2i \Delta m \varphi(t)}$, while the amplitude of the strain is slowly modulated by $Q(t)$. \par
When the decay rate is included, the perturbation starts to act roughly at time $ t_{I} \simeq -\frac{\Gamma}{2 \gamma} (1+2 z)$ \cite{extreme_mass_ratio}. If $t_{I}$ is earlier than the time when the binary enters the resonant band, the linear approximation that we employed in \cref{eqn:linear freq} no longer holds and we would have to take into account the non-linear evolution of the binary. The condition for our analysis to be self-consistent is therefore: 
\begin{figure}[t!]
    \centering
    \includegraphics[width=0.74\linewidth]{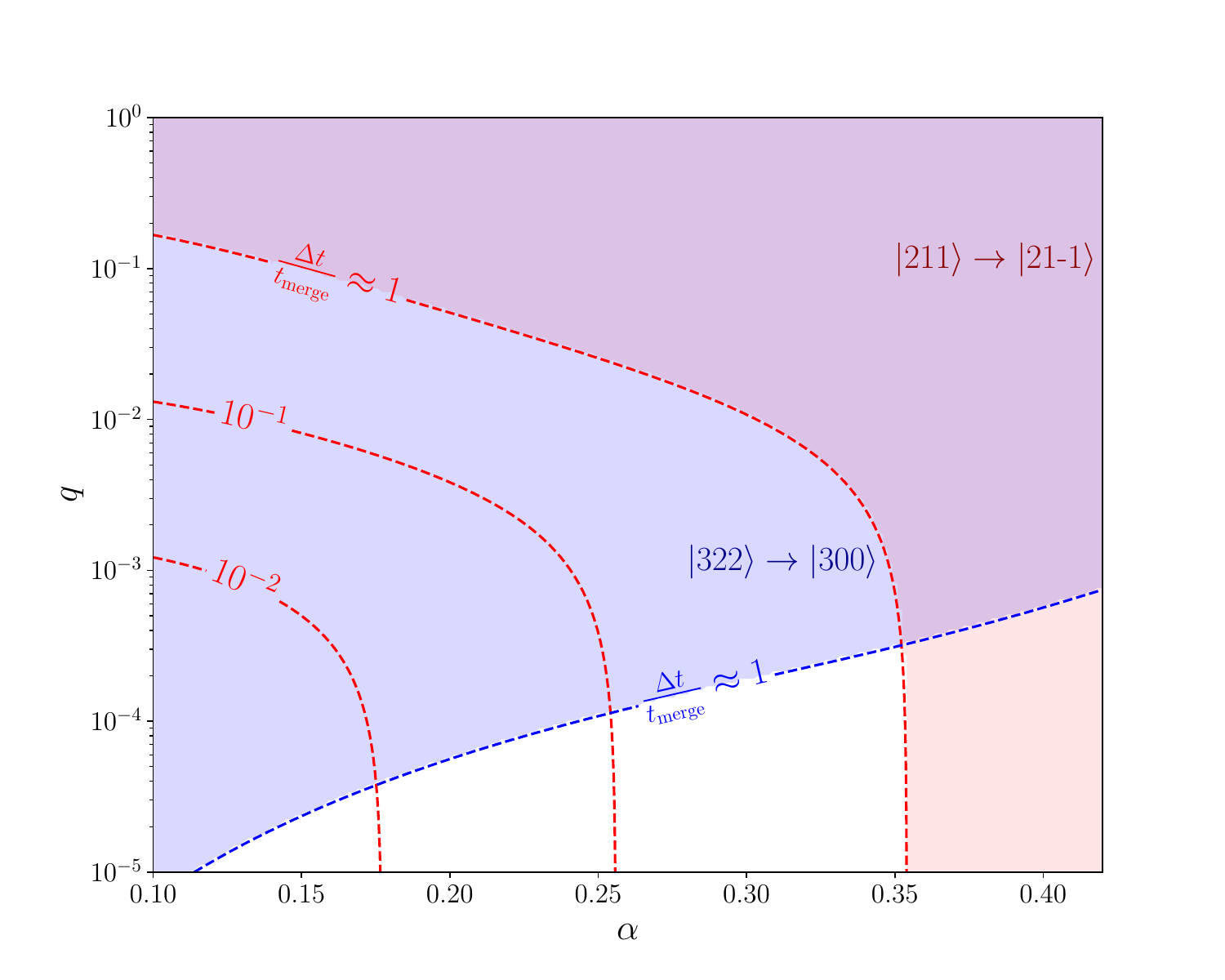}
    \caption{ Contours of $\Delta t/t_{\rm merger}$ in the $\alpha-q$ plane for two different transitions.
   }
    \label{fig:region with large timescale}
\end{figure}
\begin{equation}
    \frac{\gamma |t_{I}|}{\Omega_{0}} \simeq \frac{|\Gamma| (1+2z)}{2 \Omega_{0}} \lesssim \mathcal{O}(1).
\label{eqn:large timescale condition}\end{equation}
Equivalently, we require the duration of the transition, $\Delta t \simeq 2 |\Gamma| (1+2z)/\gamma$, to be smaller than the merger timescale when the binary is at the resonant frequency $\Omega_{0}$:
\begin{equation}
    t_{\rm merge} = \frac{5}{256} \frac{M}{(M \Omega_{0})^{8/3}} \frac{(1+q)^{1/3}}{q} \simeq 17\, \text{yrs} \left( \frac{M}{150 M_{\odot}}  \right) \left( \frac{1/150}{q} \right) \left( \frac{\alpha}{0.3}\right)^{-56/3},
\end{equation}
where $\Omega_0$ is set by $\Omega_{0,{\rm hyp}}$, in \cref{eqn:Omega hyperfine}, and it gives the same order of magnitude as given by \cref{eqn:large timescale condition}. In \cref{fig:region with large timescale}, we plot the contours of $\Delta t/t_{\rm merger}$ in an $\alpha - q$ plane for the $|211\rangle \rightarrow |21\text{-}1\rangle$ and $|322\rangle \rightarrow |300\rangle$ transitions. We have used the relativistic computation of the decay rate for the final state \cite{hoof_git,hoof2024gettingblackholesuperradiance}. The shaded regions represent the case where $\Delta t > t_{\rm merger}$. For the $|211\rangle \rightarrow |21\text{-}1\rangle$ transition, the linear approximation of the orbital frequency is not valid for $\alpha>0.35$ and all values of $q$, while for smaller $\alpha$ values, the ratio is $q$ dependent. Based on this result, we limit our attention in what follows to $\alpha < 0.35$ and mass ratios that satisfy the condition $\Delta t< t_{\rm merger}$.
\par
The fine transition $|322\rangle \rightarrow |300\rangle$ covers a large parameter space because the $|300\rangle$ state has a large decay rate compared to the $|21{\text-}1\rangle$ state. The parameter space for this transition shrinks further when the effect of the termination of superradiance due to the off-resonant mixing of the two states is considered, which excludes mass ratios that satisfy $q \gtrsim 10^{-4}$ (this affects all fine transitions while the $|211\rangle \rightarrow |21{\text-}1\rangle$ is unaffected)\,\cite{Tong_2022}. For the rest of the paper, we will only consider the $|211\rangle \rightarrow |21{\text-}1\rangle$ transition in the parameter space that was mentioned and leave the treatment of the $|322\rangle$ state for future work.
\par
In \cref{fig: strain freq time} (right), we show the frequency spectrum of the GW signal, derived using the stationary phase approximation (see details in \cref{sec:freq spectrum}), and the short-time Fourier transform (SFT) of the strain. The frequency width accessible by the Fourier transform is equal to $\gamma T_{\rm obs}$ (we have chosen an unrealistically long observation time in the figure for display purposes). We note the good agreement between the two, except for the edges of SFT results due to the windowing effect \cite{Droz:1999qx}.
The peak frequency, that is shown as a vertical dashed line, is given by (derived in \cref{sec:freq spectrum})
\begin{equation} 
    f_{\textrm{p}} = f_{\textrm{c}} - \frac{2 z |\Gamma|}{\pi}.
\label{eqn:peak freq}\end{equation}
Either in the case $z\ll 1$ or when the final-state dacay timescale is much larger than the orbital period at resonance, the peak frequency reduces to $f_{\rm p} = f_{\rm c} = 4 f_{0}$. The conclusion is that the signal is monochromatic, something that is anticipated, given that we have linearized the orbital frequency that drives the resonance around the corresponding frequency, offset however by the decay rate. In the shaded regions of \cref{fig:region with large timescale}, a signal that spans many frequency bins is expected, a possibility that we leave for future work.  \par
We may now assess the validity of the quadrupole approximation that we have used throughout. In the case where $z \ll 1$, the comparison of the wavelength of radiation to the size $r_{\rm c}$ of the system gives:

\begin{equation}
    \frac{\lambda}{r_{\rm c}}  = \frac{\alpha^{2}}{4 M f_{0}} \simeq 2 \times10^{3} \left( \frac{0.3}{\alpha}\right)^{5},
\end{equation}
and therefore the approximation is excellent. It becomes even better for larger values of $z$, where the peak frequency is smaller.
\paragraph{Comparison to inspiral and annihilation GWs}
\begin{figure}
    \centering
    \includegraphics[width=0.8\linewidth]{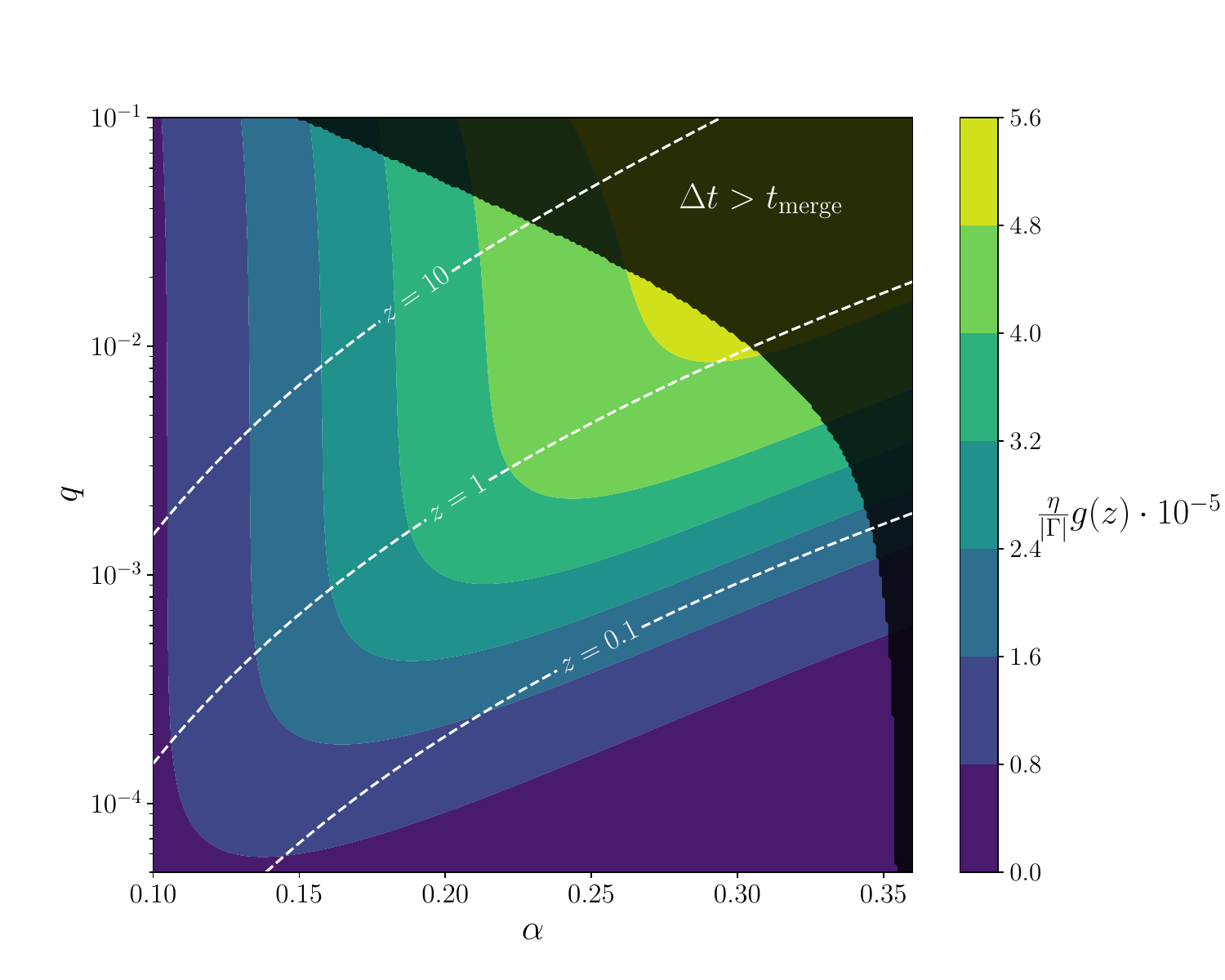}
    \caption{The factor $\frac{\eta}{|\Gamma|} g( z)$ in the $\alpha-q$ plane. Dashed lines show three values of the adiabaticity parameter $z$. The upper right corner is not considered because of the duration time being larger than the merger time.}
    \label{fig:g vs q}
\end{figure}

It is intriguing to compare the strain amplitude of the GW signal we have presented above to that of the other GW signals that originate from the inspiral itself and the annihilation of the GA. \par
Firstly, the order of magnitude of the maximum amplitude of the transition's GW signal is found by evaluating $| h_{\textrm{+}}|$ at $t_{\rm max}$ and averaging over the fast oscillations:
\begin{equation}
    h_{\textrm{max}}  = 4|\Delta m|^{2} h_{0} \frac{\eta}{|\Gamma|} g(z),
\end{equation}
where $ g( z)$ is a function of the adiabaticity parameter:
\begin{equation}
   g(z) = \frac{e^{-\pi z  + 2  z \tan^{-1}(2  z)}}{\sqrt{1+4z^{2}}},
\label{eqn: g(z)}\end{equation}.
The quantity $\eta g(z)/|\Gamma|$ is plotted in \cref{fig:g vs q} in the $\alpha-q$ plane. We notice that it is maximized for $\alpha \sim 0.3$ and $q \sim 10^{-2}$. 

The inspiral GW strain amplitude is given by \cite{Maggiore:2007ulw}, 
\begin{equation}
    h_{0,\rm b} = \frac{4}{r} \frac{q}{(1+q)^{1/3}} M^{5/3} \Omega^{2/3}_{0}.
\label{eqn:binary gw}\end{equation}
For small $ z$, the transition GW's frequency $f_{\rm p}$ is expected at twice the inspiral GW's frequency $f_{\rm insp}$ from \cref{eqn:peak freq}, i.e., $f_{\rm p}=2f_{\rm insp} = 4 f_{0}$. \par
\begin{figure}
    \centering
    \includegraphics[width=0.9\linewidth]{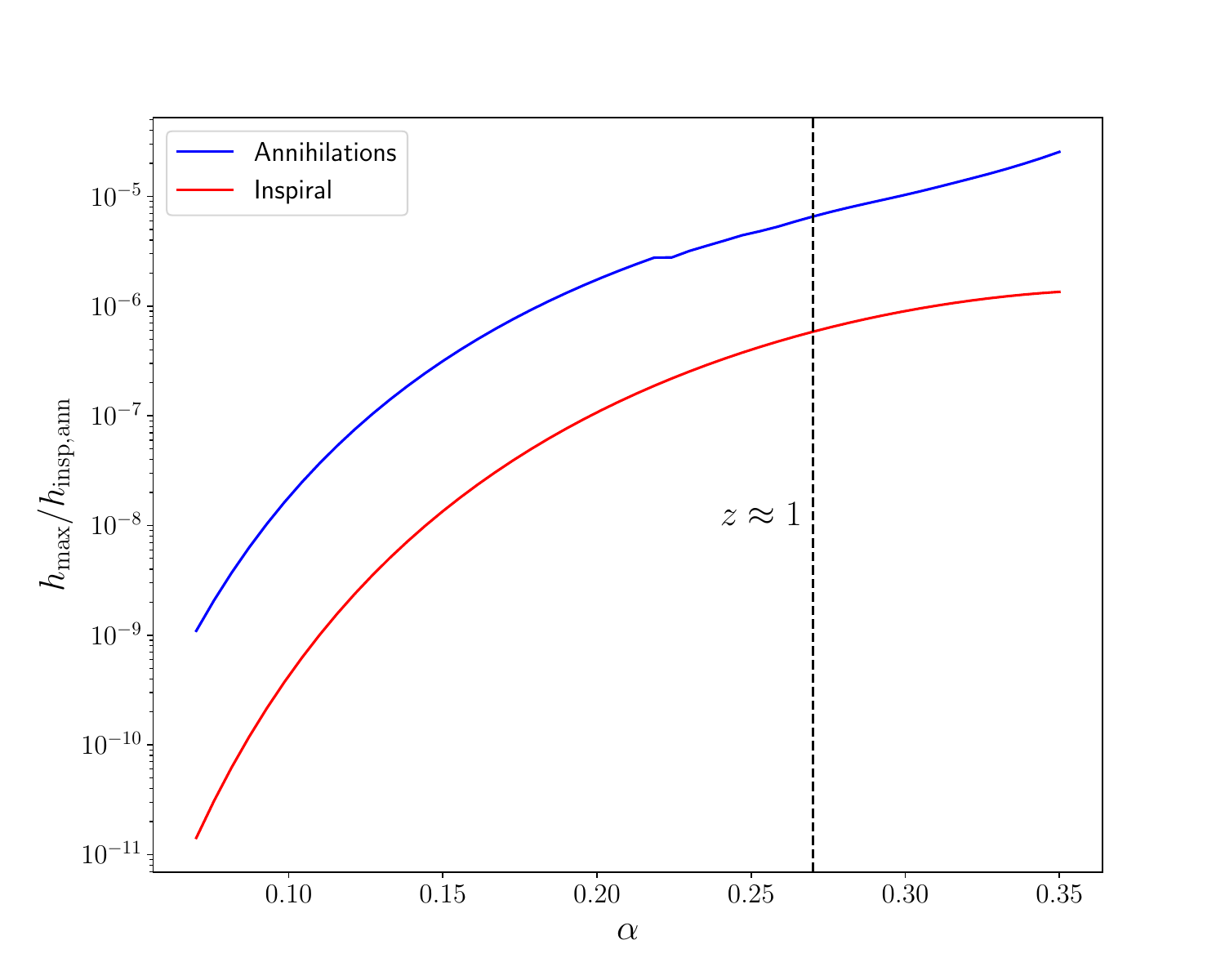}
    \caption{The ratio between $h_{\rm max}$  and the amplitudes of the GWs from the inspiral and the annihilations of the GA for $q = 1/150$ and $\tilde{a}_{\rm in}=0.99$. The vertical dashed line shows the $\alpha$ value where the adiabaticity $z \simeq 1$. For smaller $\alpha$, $z>1$ and the signal is suppressed by the exponential function in \cref{eqn: g(z)}. }
    \label{fig:amp ratio}
\end{figure}

We also consider the GWs emitted by the annihilation of the bosons in GA, with frequency equal to
\begin{equation}
    f_{\rm ann} =  \frac{\mu}{\pi} \simeq 145 \,\textrm{Hz} \left( \frac{\alpha}{0.3} \right) \left( \frac{150 M_{\odot}}{M} \right).
\label{eqn:ann frequncy}\end{equation}
The SuperRad package allows us to calculate the GW strain amplitude for given $\alpha$, $M$ and $\tilde{a}_{\rm in}$ \cite{Siemonsen:2022yyf,May:2024npn}. We calculate the ratio of $h_{\rm max}$ with these two amplitudes in \cref{fig:amp ratio} for fixed $q=1/150$ as a function of $\alpha\in[0.07,\,0.35]$. We observe that the amplitude of the transition GW signal is much smaller than the amplitudes of the other two signals. The ratio becomes very small in the small $\alpha$ limit, since $z \gg 1$ in this limit (see \cref{eqn:z hyperfine}) and the $g(z)$ function is exponentially suppressed. Varying the mass ratio within the parameter space of \cref{fig:region with large timescale} does not change this picture.    

\section{Detectability of a transition GW signal\label{section:application}}

\subsection{Relevant timescales for a transition GW signal}
The GA of initial mass $M_{c,0}$ continuously depletes its mass due to the annihilation of the bosons into gravitons.\,\footnote{This is only true for a real scalar field. A complex scalar field is expected to be long-lived \cite{Baumann_2019}.} The evolution of its mass as a function of time is given by 
\begin{equation}
    M_{c}(t) = \frac{M_{c,0}}{1+ t/\tau_{\rm ann}},
\end{equation}
where an analytical estimate for the annihilation timescale $\tau_{\rm ann}$ is given by \cite{Yoshino:2013ofa,Gravitaitonal_wave_searches,Arvanitaki_2015,LIGOScientific:2021rnv},
\begin{equation}
    \tau_{\rm ann} \simeq 9.7 \times 10^{5} \,\text{yrs} \left( \frac{M}{150 M_{\odot}} \right) \left( \frac{0.1}{\alpha}\right)^{15} \left( \frac{1}{\tilde{a}_{\rm in}} \right).
\label{eqn:annihilation timescale}\end{equation}
This formula underestimates the timescale for larger values of $\alpha$ and numerical results are needed.

The relevant timescales in our case are the annihilation timescale $\tau_{\rm ann}$, the superradiance timescale $\tau_{\rm sr}\equiv\Gamma_{211}^{-1}$, and the orbital evolution timescale during the resonant transition $\tau_{\rm orb}\equiv \Omega_0/\gamma$, assuming the orbital frequency is changed only through GW emission. If $\tau_{\rm sr}\ll \tau_{\rm ann}$, the $|211\rangle$ state can form before it completely dissipates by annihilation; otherwise, the $|211\rangle$ state cannot form. We find that the condition for the formation of $|211\rangle$ state is always satisfied in the parameter space of interest. Similarly, we require $\tau_{\rm sr}\ll\tau_{\rm orb}$ in order to build up the $|211\rangle$ state before entering the resonance band, which is also always satisfied. Furthermore, to have a long-lasting transition GW signal, we require $\tau_{\rm orb}\ll\tau_{\rm ann}$; otherwise, the $|211\rangle$ state can decay completely before the resonant transition happens. However, we numerically find that $\tau_{\rm orb}\gg\tau_{\rm ann}$ for most of the parameter space of interest, which places a strong constraint on the formation time of the $|211\rangle$ state: it needs to be formed right before the resonant transition happens in order to have detectable transition GW signals. The probability of this occurring would require a detailed study of the evolution history of binary systems given a realistic astrophysical environment and their populations, which is beyond the scope of this work. In the following, we assume the above formation time condition of the $|211\rangle$ state is satisfied.

Even though the $|211\rangle$ state can decay sufficiently fast by annihilation before the resonance takes place, there is still a possibility that higher superradiant states can build up and survive until the resonances occur. In particular, the growth timescale of the next superradiant state $|322\rangle$ is comparable to \cref{eqn:annihilation timescale} \cite{Tomaselli:2024faa}, which means that this state will have grown by the time the $|211\rangle$ state has decayed by the annihilations. We have left the treatment of transitions for higher energy superradiant states for future work, and the formalism we have developed can be straightforwardly applied to these cases as well. 
\par 
In many previous works, given these uncertainties about the annihilation of the GA and the mass of the cloud, its mass has been treated as a free parameter when performing calculations \cite{Tomaselli:2024faa,resonant_history,extreme_mass_ratio}. In what follows, we have chosen to calculate the mass of the cloud at saturation using the SuperRad package \cite{Siemonsen:2022yyf,May:2024npn} for a given boson mass, black hole mass and initial spin, with the understanding that this serves as an upper bound for the mass of the GA.   

\begin{figure}
\centering
\begin{subfigure}{.6\textwidth}
  \centering
  \includegraphics[width=0.99\linewidth]{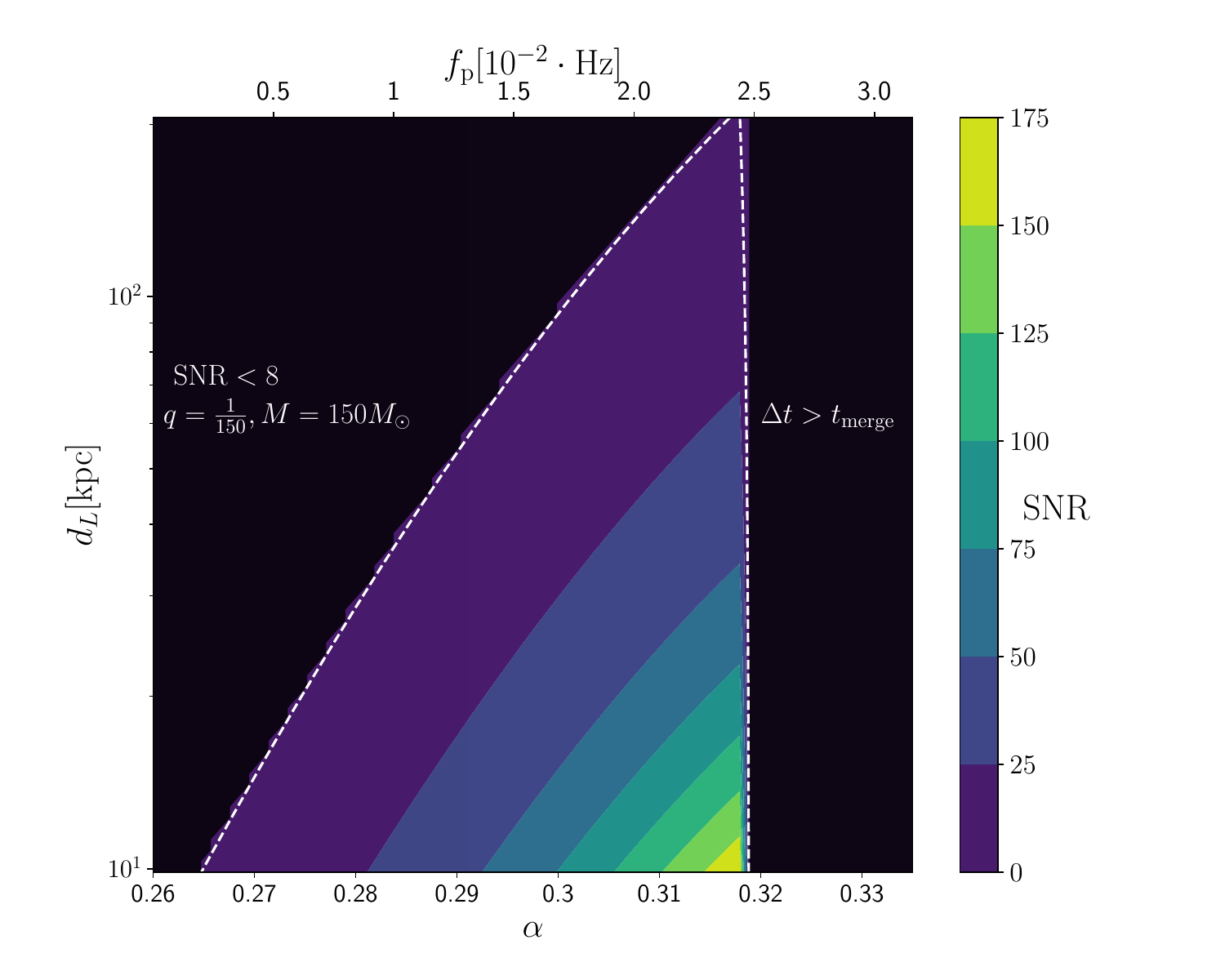}
  \end{subfigure}
  \hfill
  \begin{subfigure}{.49\textwidth}
  \centering
  \includegraphics[width=1.05\linewidth]{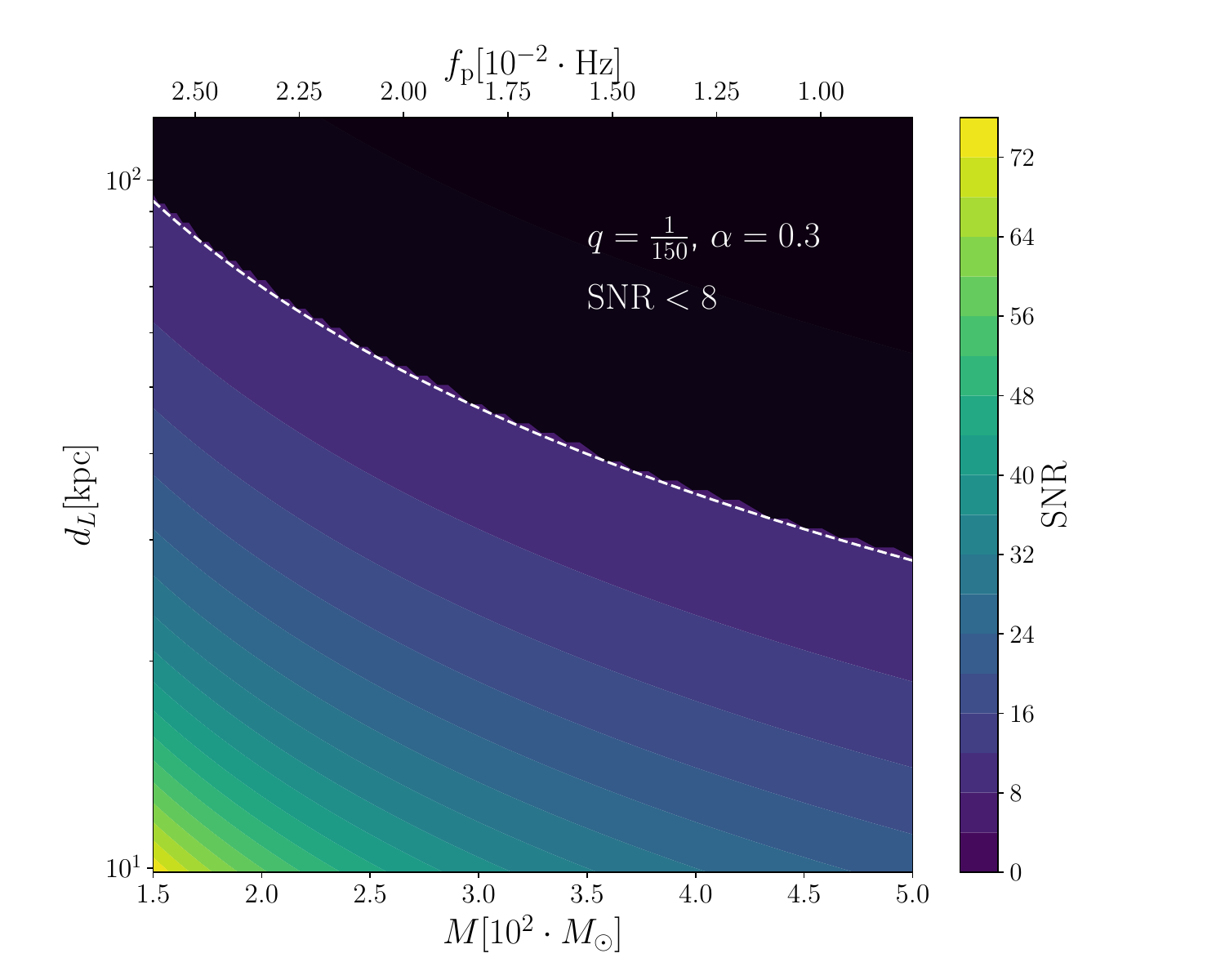}
\end{subfigure}%
 \begin{subfigure}{.49\textwidth}
  \centering
  \includegraphics[width=0.99\linewidth]{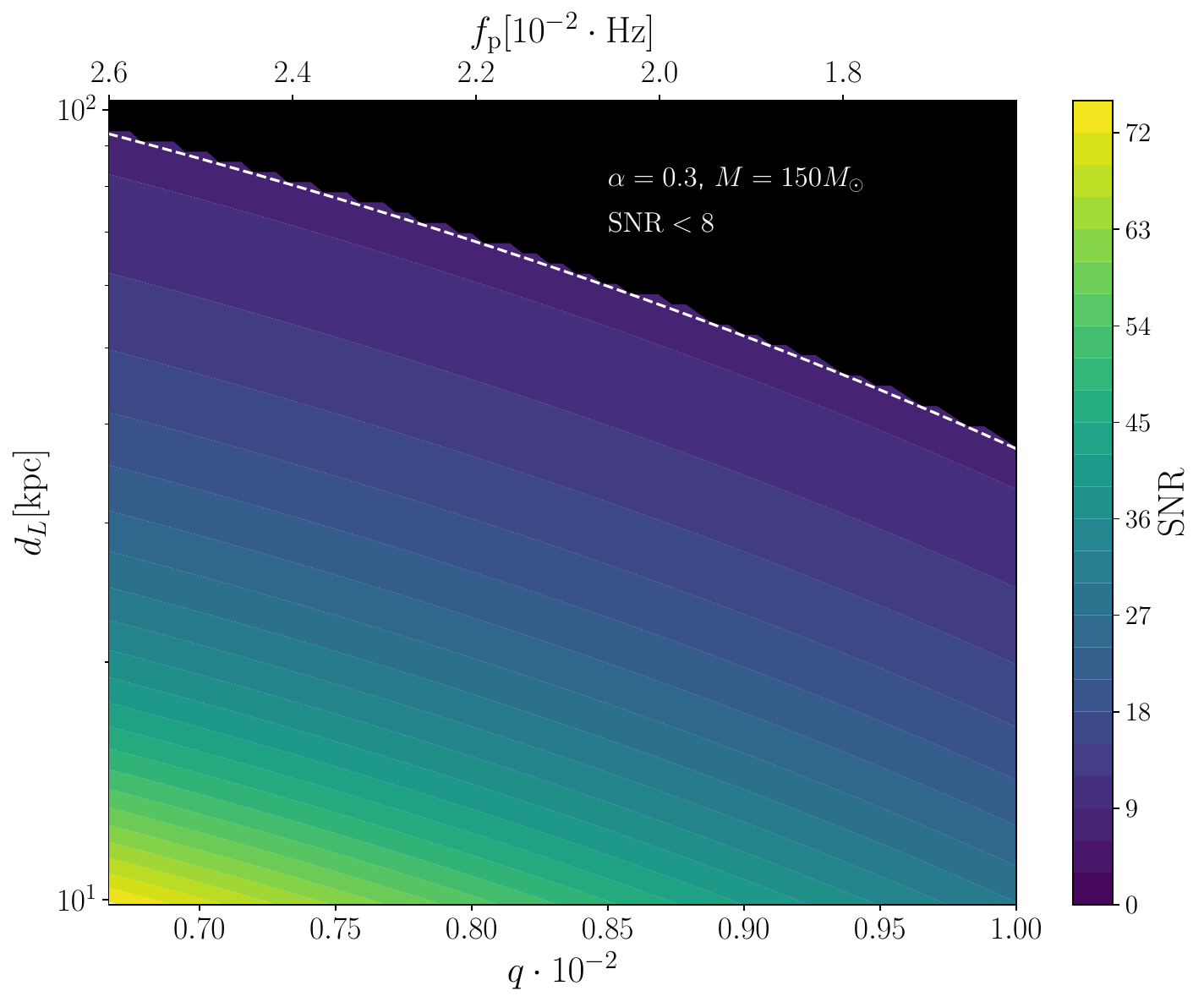}
\end{subfigure}
\caption{SNR contours, using DECIGO's sensitivity curve, on the $\alpha - d_{L}$ (top), $M-d_{L}$ (bottom left) and $q-d_{L}$ (bottom right) planes for initial black hole spin $\tilde{a}_{\rm in} = 0.99$.  }
\label{fig:d_L-params decigo 099}
\end{figure}

\begin{figure}
\centering
\begin{subfigure}{.6\textwidth}
  \centering
  \includegraphics[width=0.99\linewidth]{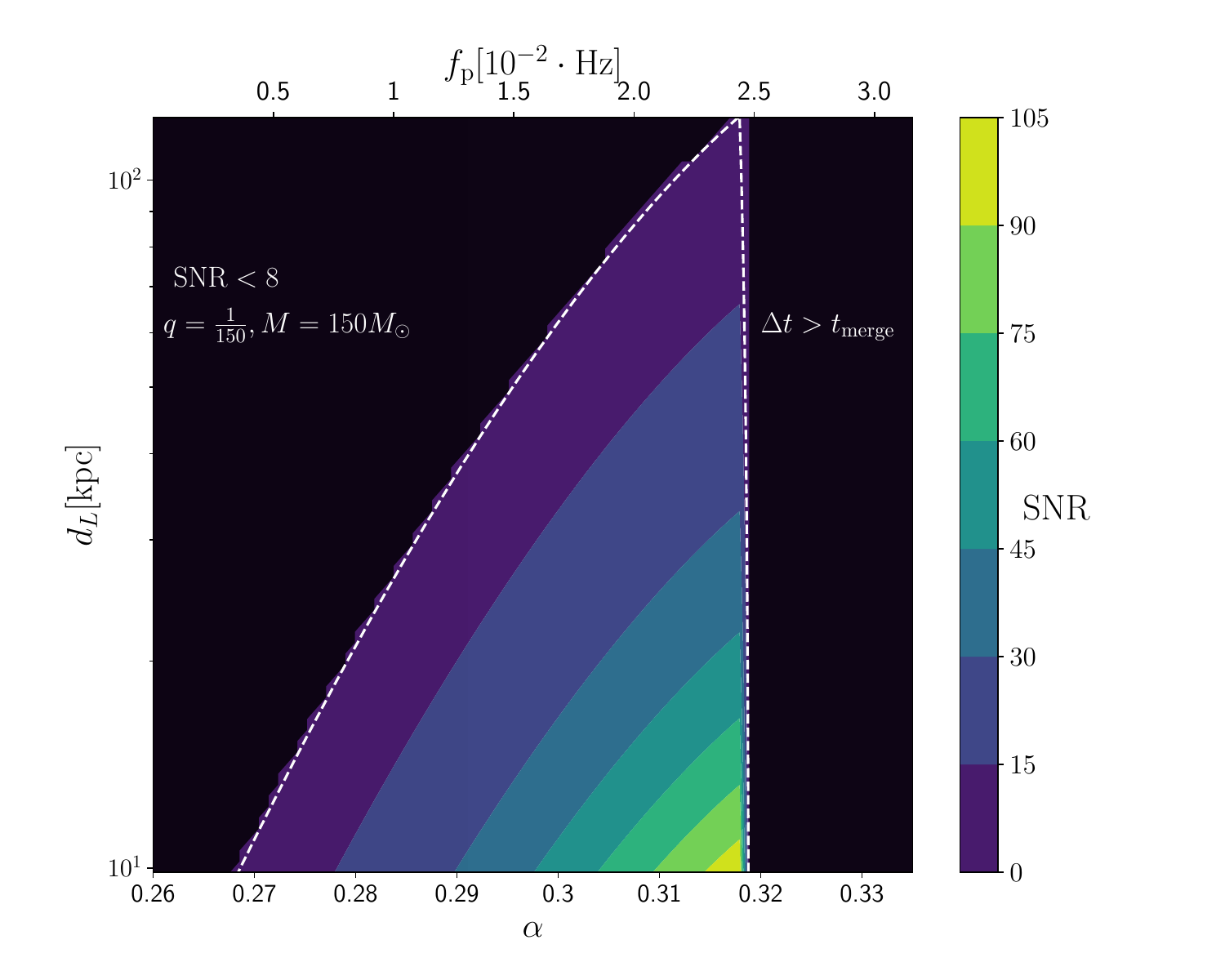}
  \end{subfigure}%
  \hfill
  \begin{subfigure}{.49\textwidth}
  \centering
  \includegraphics[width=1.05\linewidth]{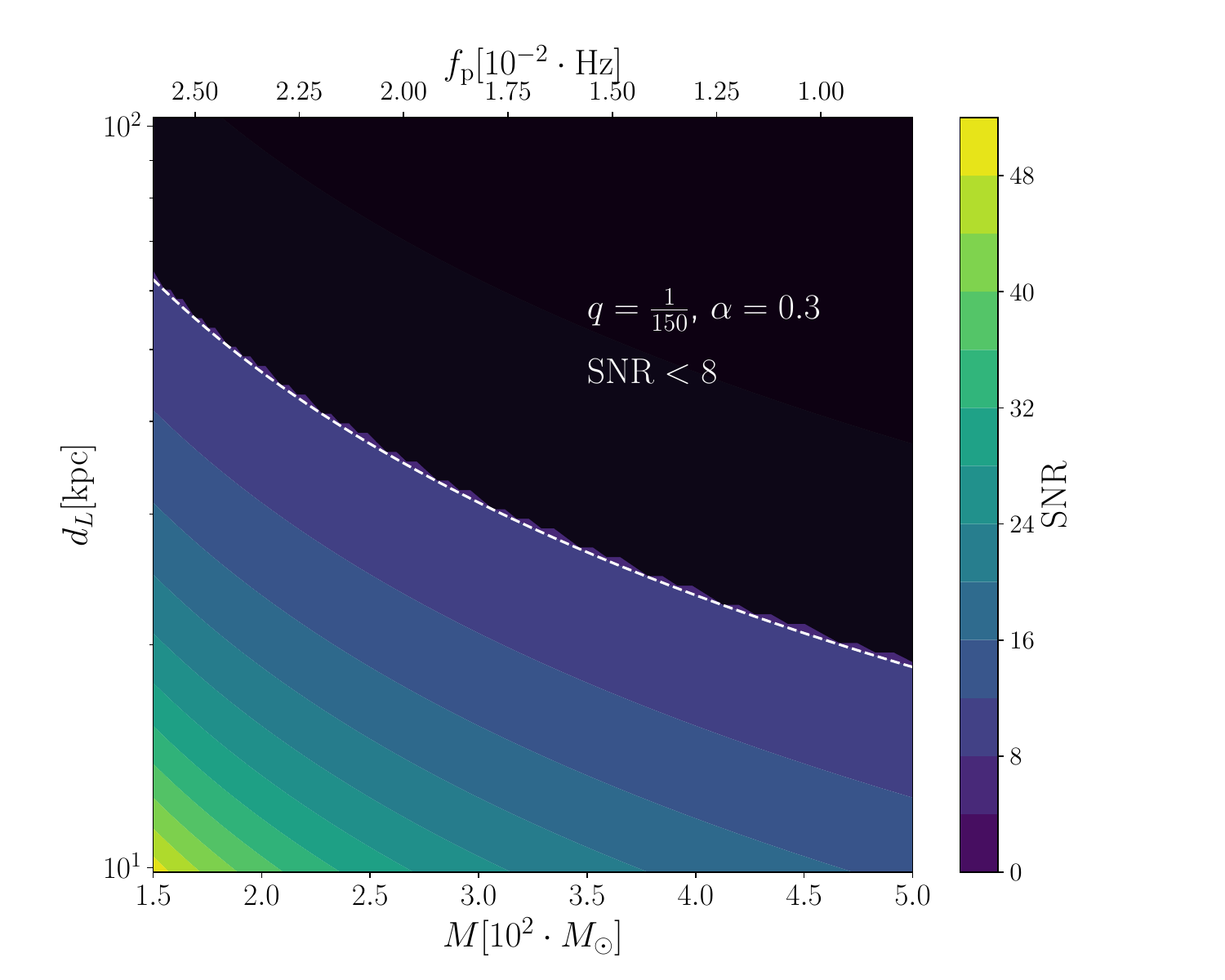}
\end{subfigure}%
 \begin{subfigure}{.49\textwidth}
  \centering
  \includegraphics[width=0.99\linewidth]{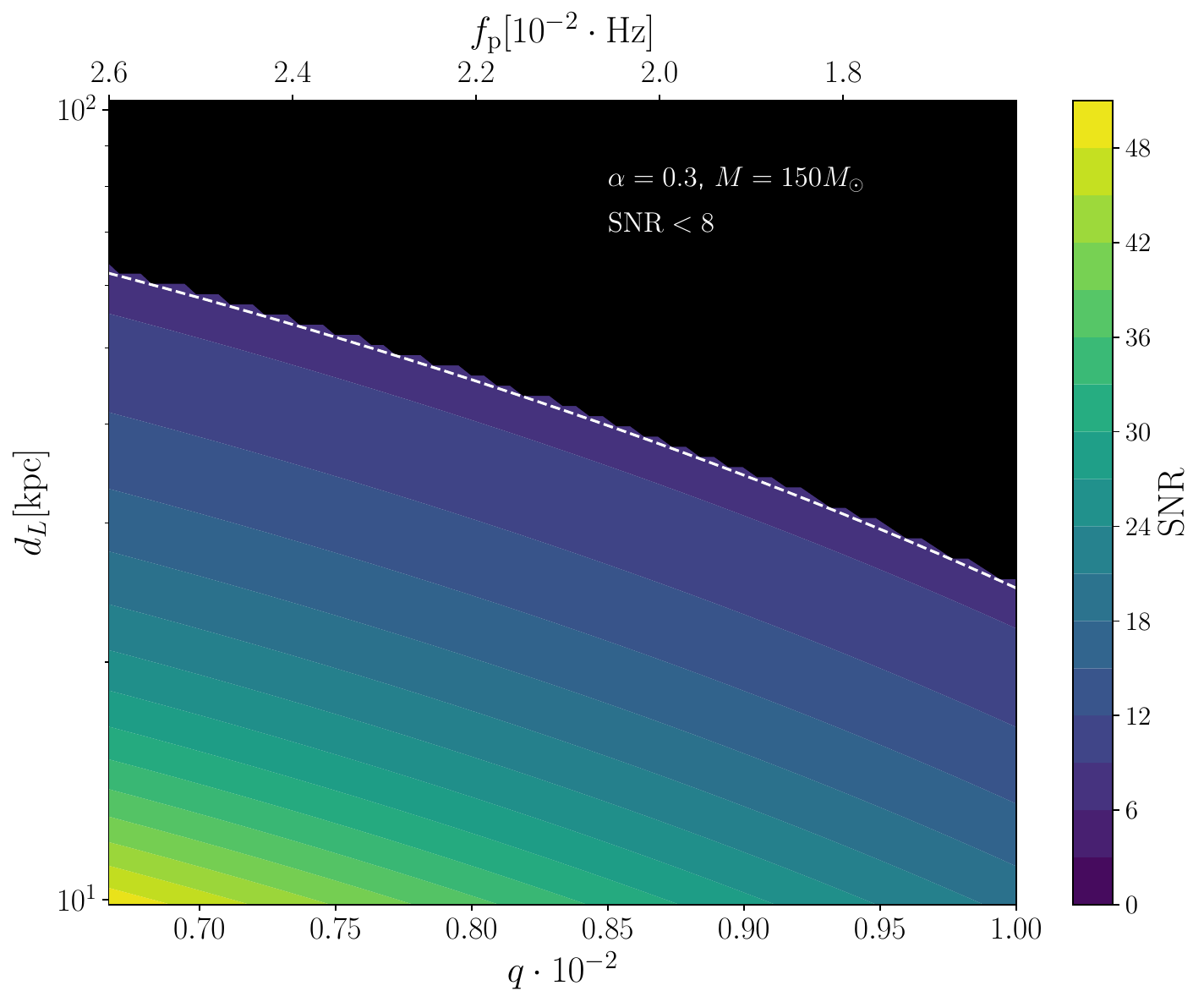}
\end{subfigure}
\caption{SNR contours, using DECIGO's sensitivity curve, on the $\alpha - d_{L}$ (top), $M-d_{L}$ (bottom left) and $q-d_{L}$ (bottom right) planes for initial black hole spin $\tilde{a}_{\rm in} = 0.95$.}
\label{fig:d_L-params decigo 095}
\end{figure} 
\subsection{The signal-to-noise ratio\label{sec:snr}}
\par 
A metric to estimate the detectability of our signal is the SNR, which for the monochromatic transition GW signal has the form \cite{LISA:2024hlh}: 
\begin{equation}
    {\rm SNR} = \mathcal{A} \frac{\eta}{|\Gamma|} h_{0} \sqrt{\frac{T_{\rm obs}}{ S_{n} ( f_{\rm p})}}  g( z) ,
\label{eqn:SNR mono}\end{equation}
where $S_{n}(f)$ is the noise power spectrum of a detector, $\mathcal{A}$ is a numerical prefactor that depends on the detector's geometry with $\mathcal{A} = \sqrt{512 \pi/5}$ for LISA and $\mathcal{A} = \sqrt{1024 \pi N_{\rm unit} / 15}$ for DECIGO, with $N_{\rm unit}$ the number of clusters in the detector \cite{DECIGO}, and $ g( z)$ is a function of the adiabaticity parameter that originates from the frequency spectrum in \cref{eqn:Ip with expo} when it is evaluated at the peak frequency \cref{eqn:peak freq} and it is given by \cref{eqn: g(z)}.
 \par
We used \cref{eqn:SNR mono} to compute the SNR for a variety of physical systems for given $\alpha$, black hole mass $M$, mass ratio $q$, initial black hole spin $\tilde{a}_{\rm in}$, and luminosity distance $ d_{ L}$\,\cite{Maggiore:2007ulw}.  We numerically determined the mass of the cloud at saturation using the SuperRad module\,\cite{May:2024npn,Siemonsen:2022yyf}, as discussed in the previous subsection. The decay rate of the second state was calculated numerically for all cases using the code in \cite{hoof2024gettingblackholesuperradiance,hoof_git} and assuming again that the black hole's spin has reached its saturated value. We used $ T_{\rm obs} = 4  \rm yrs$ for all calculations.
\par
Given the allowed parameter space for our analysis shown in \cref{fig:region with large timescale}, we have focused on intermediate mass ratio binaries $q \sim 10^{-3} - 10^{-2}$ and intermediate-mass black holes (IMBHs) $M \geq 150 M_{\odot} $, which may be the massive host of such binary systems in dense nuclear and globular clusters \cite{MacLeod:2015bpa,askar2024intermediatemassblackholesstar}. In particular, if they form through successive mergers of lower mass black holes, they can also have large spins, making them ideal for the superradiance process \cite{Berti:2008af,Borchers:2025sid}. In fact, the LIGO event GW231123 has recently confirmed a discovery of a binary system with a black hole remnant mass of $190-265M_\odot$, resulting from a merger of two $\sim 100 M_\odot$ black holes with high spins $\sim 0.8-0.9$ \cite{LIGOScientific:2025rsn}. Regarding the detection prospects in future GW experiments, LISA's  is expected to detect $0.02-60$ such events per year, while for DECIGO that number is $6-3000$ per year, up to red-shift $z \approx 10$ (Figure 20 of \cite{Arca-Sedda:2020lso}). There are also hints that point to the existence of IMBHs in globular clusters in the Milky Way and in its satellite galaxies, though there is ongoing debate in the literature surrounding these findings \cite{10.1093/mnras/stw1779,Kiziltan:2017ijz,Abbate_2019, 2019MNRAS.487.2685E,2017ApJ...846...14B,2007MNRAS.376L..29G}. \par

In \cref{fig:d_L-params decigo 099}, we plot the $\rm SNR$ contours for the DECIGO telescope in the $\alpha-d_{L}$, $M-d_{L}$ and $q - d_{L}$ planes for an initially maximally spinning black hole. In general, we were only able to find a detectable signal for systems within $100 \,\rm kpc$. In the $\alpha -  d_{ L}$ plot, we observe that the $\rm SNR$ increases as $\alpha$ increases, due to the strong dependence of GW strain amplitude on $\alpha$ (\cref{eqn:amp-scaling}) and the fact that the peak frequency $f_{\rm p}=4f_0-2z|\Gamma|/\pi$ reaches the most sensitive frequency range of DECIGO as $\alpha $ increases. In our analysis, we cannot increase $\alpha$ past $\alpha \sim 0.32$ since in that case the duration of the signal would be comparable to the merger timescale for these parameters (see \cref{fig:region with large timescale}).   
In the $ M- d_{ L}$ plot, we note that the strongest $\rm SNR$ is produced by the smallest black hole masses, since larger masses imply smaller peak frequency, outside of the experiment's reach. Finally, in the $ q -  d_{ L}$ plot, we observe that the low mass ratios are favored for the particular choice of $\alpha$ and black hole masses.
 \par
In \cref{fig:d_L-params decigo 095}, we show the same plots for an initially smaller black hole spin, $\tilde{a}_{\rm in} = 0.95$, keeping all the other parameters fixed. This results in a smaller mass for the GA, hence reducing the strength of the signal. Even smaller spins would restrict us to $\alpha<0.3$ (see \cref{fig:Superrad rate}), further reducing the $\rm SNR$. 
\par
Regarding the LISA experiment, we did not find detectable SNR values for the benchmark parameters we considered above and chose not to report those results.\par
Overall, our results are not as promising as those reported in \cite{Peng:2025zca}. The combined effects of the decay rate, which suppresses the $\rm SNR$ by the factor $\frac{\eta}{|\Gamma|}$, the initial black hole spin, which for smaller values produces a less massive GA, and the annihilation of the GA into gravitons would make this a challenging signal to observe, even for binary systems close to the Earth, for which the merger rate is expected to be very small \cite{Cheung:2025grp,Arca-Sedda:2020lso}.   A more complete study that accounts for all these effects simultaneously is left for the future. Finally, in \cref{sec:relativistic}, we establish where the non-relativistic approximation that we have employed throughout fails significantly in comparison to the full relativistic result. The biggest discrepancy is observed in the decay rate, which deviates by an order of magnitude and would lead to a larger $\rm SNR$ if the less accurate non-relativistic result is used. This establishes the importance of accounting for relativistic effects in assessing the detectability of this signal. \par
Given the effectively monochromatic nature of the GW signals from level transitions (within the parameter space considered), even if such a signal were detected, it could be degenerate with other quasi-monochromatic astrophysical sources. It is worth further studies in combining the inspiral GW signals and transition GW signals to break the degeneracy.

\section{Conclusion}\label{section:conclusion}
In this paper, we have developed a formalism to calculate the GW signal produced by the resonant transition of a GA in a binary system. In the perturbative regime where the companion's orbital distance is much larger than the size of the GA, the dominant multipole moment of the tidal field from the companion is the quadrupole moment $l_*=2$, which gravitationally mixes two GA states and thus triggers the transition.
\par
The two-state transition is modeled by the Landau-Zener system, which is characterized by the adiabaticity parameter that depends on the fine-structure constant $\alpha$, the mass ratio of the companion $q$, and the decay rate of the bosons in the final state back to the black hole. Under the assumption of a quasi-circular and equatorial orbit, the analytical formulae of the GW strain waveform and frequency spectrum are derived in the linearized orbital frequency evolution region. We validate the parameter region of $(\alpha,\,q)$ where the timescale of the transition is less than the merger time of the binary for the $|211\rangle \rightarrow |21 \text{-}1\rangle$ hyperfine and $|322\rangle \rightarrow |300\rangle$ fine transitions. In the present work, we restrict our attention to the $|211\rangle \rightarrow |21 \text{-}1\rangle$ hyperfine transition for mass ratios $q \lesssim 10^{-2} $ and $\alpha < 0.35$.    \par
The GW strain waveform can be considered as a continuous signal with amplitude modulated by the LZ transition. Its analytical formula is given in \cref{eqn: plus strain} and it is plotted in \cref{fig: strain freq time} for the benchmark parameters. We stress that the analytical waveform of the signal can serve as the template used by the match-filtering technique for future GW searches. The frequency spectrum of the strain is also derived analytically using the stationary phase approximation and its peak frequency is derived in terms of the fundamental parameters in \cref{eqn:peak freq} and it is plotted in \cref{fig: strain freq time}. In the parameter space of interest, the transition GW signal is always monochromatic.
\par
In terms of the detectability of the signal, we calculated the SNR for a variety of systems and determined that the DECIGO experiment would be suited for detecting it from systems that host IMBHs with $M \sim 150 M_{\odot}$, and $q \sim  1/150$, with $\alpha \sim 0.3$, which corresponds to boson masses $10^{-13} \rm eV$. The maximum distance that the signal can be detected at is $100 \,\rm kpc$. We also explored the possibility of a smaller initial black hole spin and found that it reduces the strength of the signal, as it implies a smaller mass for the GA.
\par
Finally, we commented on the validity of the non-relativistic approximation. We determined that the relativistic radial wavefunction is well approximated by its non-relativistic counterpart, even for moderately large values of $\alpha$, and the same holds true for the energy levels of the GA. We spotted the biggest discrepancy in the decay rate of the second state, where the relativistic result deviates up to an order of magnitude from the non-relativistic approximation for large $\alpha$. Given how sensitive the signal is to this decay rate, we high-lighted the importance of including the accurate result in the calculations.
\par
To streamline the analysis, several assumptions were made. We now discuss potential follow-up studies in the future to improve it:
\begin{itemize}
    \item An important continuation of the present study would be to extend its regime of validity to the entirety of the parameter space of \cref{fig:region with large timescale}, by including the non-linear evolution of the orbit throughout the transition.
    \item The simplifying assumption of 
quasicircular and equatorial orbits could be refined, especially for extreme mass ratio inspirals, which are expected to be highly eccentric. It has been found that if we relax these assumptions, transitions can occur at a variety of frequencies \cite{resonant_history,Berti:2019wnn}. The detectability of the GA transition signal in this case is something worth examining.
\item Taking into account the back-reaction of the transition to the orbit is another important topic. This is manifested through the conservation of angular momentum for the entire system.
The GWs that we have found will contribute to the total torque of the system, possibly leading to non-trivial dynamics of the orbit. This interplay between the back-reaction of the transition and the back-reaction of the emitted GWs might lead to unexpected and interesting behavior for the binary's orbit.
\item In terms of environmental effects, since the black holes that would produce the largest $\rm SNR$ are of intermediate mass that are typically found in dense globular clusters, incorporating the perturbations to the GA from nearby stars and quantifying their influence on the GA's evolution, perhaps in the spirit of \cite{Dandoy:2022prp}, would be an interesting direction.
\item Finally, a more complete treatment of the relativistic GA \cite{witte2025steppingsuperradianceconstraintsaxions}, the perturbations from the companion \cite{Li:2025ffh} and the effects those would have on the signal, would be an important addition to the present study.
\end{itemize}

\acknowledgments
We are grateful to Hyeonseok Seong for many interesting discussions. We are also thankful to Astrid Lamberts for providing the FIRE simulation dataset. Additionally, we want to thank Jeff Dror, John Stout, Rachel Houtz, Wei Xue, Yue Zhao and Joseph Fichera for their insightful and useful discussions. F.Y. and A.K ~are supported in part by the U.S.~Department of Energy under grant DE-SC0022148 at the University of Florida. AK is also supported by the Onassis Foundation - Scholarship
ID: F ZS 031-1/2022-2023.

\appendix
\section{Gravitational atom basics}
\subsection{Tidal field of the companion}
\label{appendix: tidal}
We present in this appendix some of the formulae we used for the tidal field of the companion and the Hamiltonian for the two state system.
The tidal field is given as a sum over multipoles:

\begin{equation}
    V_{\ast}(t,\vec{r}) = -q \alpha \sum_{l_{\ast} = 2} \sum_{|m_{\ast}|\leq l_{\ast}} \mathcal{E}_{l_{\ast},m_{\ast}}(\iota_{\ast},\varphi_{\ast}) Y_{l_{\ast},m_{\ast}}(\theta,\phi) \left(\frac{r^{l_{\ast}}}{R^{l_{\ast}+1}} \Theta(R_{\ast} - r) + \frac{R_{\ast}^{l_{\ast}}}{r^{l_{\ast}+1}} \Theta(r - R_{\ast})\right),
\end{equation}
where $R_{\ast}$ is the distance of the companion from the primary black hole, $\mathcal{E}_{l_{\ast},m_{\ast}} (\iota_{\ast},\varphi_{\ast})$ is the tidal moment (see Appendix A of \cite{Baumann_2020} for details), $\iota_{\ast}$ is the inclination of the orbital plane relative to the GA's, $\varphi_{\ast}$ is the true anomaly and is related to the frequency of the binary via $\dot{\varphi_*}(t) = \pm \Omega(t) $. We have also defined $q = \frac{M_{\ast}}{M}$ as the mass ratio.

Assuming that the quadrupole dominates, the mixing between two states $| \psi_{i} \rangle$ and $| \psi_{f} \rangle$ is given by
\begin{equation}
   \langle \psi_{f} |V_{\ast} (t,\vec{r}) | \psi_{i} \rangle  = \mathcal{E}_{2,m_{\ast}} (\iota_{\star},\varphi_{\ast}) I_{A} \left( \frac{ q M \Omega }{\alpha^{3} (1+q)} I_{\textrm{in}} + \frac{\alpha^{7} q (1+q)^{2/3}}{(M \Omega)^{7/3}} I_{\textrm{out}} \right),                            
\label{eqn:V}\end{equation}
where the inner radial, outer radial, and angular overlaps of the initial and final wavefunction mediated by the quadrupole moment of the tidal field are:
\begin{align}
    & I_{\textrm{in}}  = \int^{R_{\ast}/r_{c}}_{0} d \bar{r} \bar{r}^{4} \bar{R}_{i} (r_{c} \bar{r}) \bar{R}_{f} (r_{c} \bar{r}),  \\
    & I_{\textrm{out}} = \int^{\infty}_{R_{\ast}/r_{c}} d \bar{r} \bar{r}^{-1} \bar{R}_{i} (r_{c} \bar{r}) \bar{R}_{f} (r_{c} \bar{r}), \\
    & I_{A} = \int d \Omega Y_{l_{\ast},m_{\ast}} (\theta,\varphi) Y_{i}(\theta,\varphi) Y^{\ast}_{f}(\theta,\varphi), 
\end{align}
with new dimensionless quantities $\bar{r} = \frac{r}{r_{c}}$ and $\bar{R} (r_{c} \bar{r})=R(r)r^{3/2}_{c}$. The selection rules \cref{eq:selection_rule} are derived from the last angular integral. Taking $\iota_{\ast} = 0$, the amplitude of the mixing normalized by the orbital frequency is:
\begin{equation}
    \frac{\eta}{\Omega}  = \left|\sqrt{\frac{3 \pi}{10}}  I_{A} \left( \frac{ q M \Omega }{\alpha^{3} (1+q)} I_{\textrm{in}} + \frac{\alpha^{7} q (1+q)^{2/3}}{(M \Omega)^{7/3}} I_{\textrm{out}} \right)\right|,
\label{eqn:eta}\end{equation}
and the matrix element is given in \cref{eqn:matrix element}.

\subsection{Dressed frame Hamiltonian}
\label{appendix:Hamiltonian}
\par
In the dressed co-rotating frame, the Hamiltonian is given by:
\begin{equation}
\bar{\mathcal{H}} = 
    \begin{pmatrix}
  - \frac{\Delta m}{2} (\frac{\Delta \omega_{R}}{\Delta m}  \mp\Omega(t))   &  \eta  \\
    \eta   &   \frac{\Delta m}{2} (\frac{\Delta \omega_{R}}{\Delta m} \mp \Omega(t) ) - i |\Gamma| 
\end{pmatrix}.
\label{eqn:dressed Hamiltonian}\end{equation}

Choosing $\Omega_{0} = \pm \frac{\Delta \omega_{R}}{\Delta m}$ and using \cref{eqn:linear freq}, we obtain the Hamiltonian of the Landau-Zener system (for a co-rotating orbit),
\begin{equation}
\bar{\mathcal{H}} = 
    \begin{pmatrix}
   \frac{\Delta m}{2} \gamma t   &  \eta  \\
    \eta   &  - \frac{\Delta m}{2} \gamma t - i |\Gamma|
\end{pmatrix}.
\label{eqn:Hamiltonian}\end{equation}

The second order differential equations for $d_{1}(t)$ and $d_{2}(t)$ are
\begin{align}
   & d''_{i}(\tau) + d_{i}(\tau)\left( \frac{\tau^{2}}{4}  -  \frac{i}{2}  + z \right) = 0,  \\
   & d''_{f}(\tau) + d_{f}(\tau)\left( \frac{\tau^{2}}{4}  +  \frac{i}{2}  + z \right) = 0 , 
\label{eqn:equation of motion}\end{align}
where we have defined $\tau = \sqrt{|\Delta m| \gamma} t$ and set $\Gamma =0$. The prime denotes the derivative with respect to $\tau$. The solutions to these equations are given in \cref{eq:coefficient_d1,eq:coefficient_d2} and are plotted in \cref{fig:hyperfine}.

\section{Derivation of quadrupole GW radiation}
\label{sec:quadrupole}
In this appendix, we demonstrate the derivation of the quadrupolar radiation formula for GW emission, in the limit where the wavelength of the radiation is much larger than the size of the source. Our discussion follows closely that of \cite{Maggiore:2007ulw}.\par
The GW strain outside of a source in the transverse-traceless (TT) gauge is given in \cite{Maggiore:2007ulw},
\begin{equation}
 \label{eq:h_general}
   h^{\rm TT}_{ij}({\bf x},t)=4\Lambda_{ij,lm}(\hat{\bf n})\int \frac{d^3{\bf x}'}{|{\bf x}-{\bf x}'|}T_{lm}({\bf x}',t-|{\bf x-x}'|),
\end{equation}
where $T_{lm}$ is the stress tensor of the source, $\hat{\bf n}$ is the propagation direction of GW, and $\Lambda_{ij,lm}(\hat{\bf n})\equiv P_{il}P_{jm}-\frac{1}{2}P_{ij}P_{lm}$, $i,\,j,\,l,\,m=1,\,2,\,3$, with $P_{ij}(\hat{\bf n})=\delta_{ij}-n_in_j$ the transverse projector. In the far zone limit, $r\equiv|{\bf x}|\gg|{\bf x}'|_{\rm max}$, the strain reduces to,
\begin{equation}
\label{eq:hTT_far_zone}
    h^{\rm TT}_{ij}({\bf x},t)\simeq\frac{4}{r}\Lambda_{ij,lm}(\hat{\bf n})\int d^3{\bf x}'T_{lm}({\bf x}',t-r+\hat{\bf x}\cdot{\bf x}').
\end{equation}
It is convenient to rewrite the above integral in terms of the Fourier transform of the stress tensor,
\begin{equation}
   \label{eq:T_k}
     T_{lm}({\bf x}',t-r+\hat{\bf x}\cdot{\bf x}')=\int \frac{d^4 k}{(2\pi)^4}\tilde{T}_{lm}({\bf k},\omega)e^{-i\omega(t-r+\hat{\bf x}\cdot{\bf x}')+i{\bf k}\cdot {\bf x}'},
\end{equation}
where the wavevector and frequency of the GW satisfy $|{\bf k}|=\omega$.
If $\tilde{T}_{lm}({\bf k},\omega)$ of the source peaks around a characteristic frequency $\omega_c$, with its inverse much larger than the size of the GW source $\lambda=2\pi/\omega_c\gg |{\bf x}'|_{\rm max}$, then it is valid to do the Taylor expansion of the exponent in \cref{eq:T_k}. 
As we show in \cref{eqn:peak freq}, the GW wavelength associated with the hyperfine transition is much larger than the size of the GA, $\lambda\simeq \pi/(\Delta m\Omega_{0,{\rm hyp}})\gg r_c$. So, the exponent in \cref{eq:T_k} takes the form
\begin{equation}
    e^{-i\omega(t-r+\hat{\bf x}\cdot{\bf x}')}= e^{-i\omega(t-r)}[1-i\omega \hat{\bf x}\cdot{\bf x}'+\mathcal{O}(\omega |{\bf x}'|_{\rm max})^2],
\end{equation}
which, in the leading order, is equivalent to rewriting the stress tensor
\begin{equation}
\label{eq:Tlm_far_zone}
    T_{lm}({\bf x}',t-r+\hat{\bf x}\cdot{\bf x}')\simeq T_{lm}({\bf x}',t-r)+(\hat{\bf x}\cdot{\bf x}')\partial_0T_{lm}({\bf x}',t-r).
\end{equation}
In the linearized theory, by neglecting the backreaction of the GW, the energy-momentum tensor $T_{\mu\nu}$ satisfies the conservation equation
\begin{equation}
\label{eq:conservation}
    \partial_\mu T^{\mu\nu}=0, ~~\mu,\nu=0,1,2,3.
\end{equation}
The temporal component of the conservation equation indicates mass conservation, 
\begin{equation}
    \dot{M}=\int_V d^3{\bf x}' \partial_0T^{00}=-\int_V d^3{\bf x}' \partial_iT^{i0}=-\int_{\partial V} dS^i T^{i0}=0, 
\end{equation}
where the boundary $\partial V$ is chosen outside of the source, while the spatial component indicates momentum conservation,
\begin{equation}
    \dot{P}^i= \int_V d^3{\bf x}' \partial_0T^{0i}=-\int_{\partial V} dS^j T^{ji}=0.
\end{equation}
The higher order momenta of $M$ and $P^i$ and their time derivatives can be derived similarly, and the momenta of the stress tensor $S^{ij}$ can be expressed in terms of these quantities, 
\begin{equation}
\label{eq:Sij}
    S^{ij}\equiv \int d^3{\bf x}'T^{ij}=\frac{1}{2}\ddot{M}^{ij},
\end{equation}
where $M^{ij}\equiv \int d^3{\bf x}'{x'}^i{x'}^j T^{00} (t,\vec{x}')$ is the mass quadrupole moment. Therefore, in the leading order, the GW strain is given by combining \cref{eq:hTT_far_zone,eq:Tlm_far_zone,eq:Sij} and is the result of \cref{eqn:quadrupole formula h}.

The above derivation shows that the leading contribution to the GW strain is the quadrupole radiation when the characteristic wavelength of the Fourier mode $\tilde{T}_{ij}$ is much larger than the size of the source and energy-momentum conservation is satisfied.

\section{Frequency spectrum and stationary phase approximation}
\label{sec:freq spectrum}
In this appendix, we go into more detail about the calculation of the frequency spectrum. It is given by the Fourier transform of the strain $h(t)$:
\begin{equation}
    \tilde{h}_{+,\times}(\omega) = \int dt h_{+,\times}(t) e^{i \omega t},
\label{eqn:fourier h}\end{equation}
Focusing on the plus polarization, 
\begin{equation}
    \tilde{h}_{+}(\omega)  = h_{0} \frac{1+\cos^{2}\iota}{2} \frac{e^{i \omega r }}{2}\int d t \left( e^{i(\omega t - 2 \Delta m \varphi(t))}Q(t) + e^{i(\omega t + 2 \Delta m \varphi(t))}Q^{\ast}(t) \right),
\label{eqn:Ip}\end{equation}
where we have renamed $t_{\rm re} \rightarrow t$.\par
For positive frequencies, the stationary phase approximation can be used to evaluate the above integrals. The approximation is based on the fact that the integrals are dominated at the stationary points, that is, the points where the first derivative of the phase is zero. Since $\dot{\varphi}(t) = \Omega(t)>0$, the first integral does not have a stationary point and the fast oscillations average to zero. The stationary point of the second integral is $t_{+}(\omega) = \frac{\omega - \omega_{\rm c}}{2 |\Delta m| \gamma}$, where we defined $\omega_{\rm c} \equiv 2 |\Delta m| \Omega_{0} = 4 \Omega_{0}$.
 \par
Evaluating the $Q$ factor at this stationary point, Taylor expanding the phase around the it to order $t^{2}$, and calculating the Gaussian integrals, we find
\begin{equation}
    \tilde{h}_{+}(\omega)  = \frac{\sqrt{\pi} h_{0} (1+\cos^{2}\iota)}{4 \sqrt{\gamma |\Delta m|}} Q^{\ast}(t_{+}(\omega))e^{i \Psi_{+}(\omega)}, 
\label{eqn:I factors}\end{equation}
with $\Psi_{+}(\omega) = \omega r + \frac{(\omega - \omega_{\rm c})^{2}}{4 |\Delta m| \gamma }-\frac{\pi}{4}$. Using \cref{eqn: d1^2 gamma>> 1/dt} and \cref{eqn: d2 gamma>> 1/dt}, we get
\begin{equation}
    \tilde{h}_{+}(f) =  h_{0} (1+\cos^{2} \iota) \sqrt{\pi} |\Delta m|^{2} i e^{i\Psi_{+}(f)} \frac{\sqrt{z}}{|\Gamma| - i \pi (f - f_{\rm c})} e^{- \pi z} e^{-2 z \tan^{-1}( \frac{\pi(f - f_{\rm c})}{ |\Gamma|})},
\label{eqn:Ip with expo}\end{equation}
where we re-expressed the frequency in terms of $f = \frac{\omega}{2 \pi}$. Similar results hold for $\tilde{h}_{\times}(f)$ that only differs by a phase of $\frac{\pi}{2}$. The peak frequency of $ |\tilde{h}(f)|$ is easily found by taking a derivative of \cref{eqn:Ip with expo} with respect to $f$ and setting it equal to zero, which yields \cref{eqn:peak freq}.
\section{Towards a relativistic computation}
\label{sec:relativistic}
\begin{figure}[h!]
    \centering
    \includegraphics[width=0.7\linewidth]{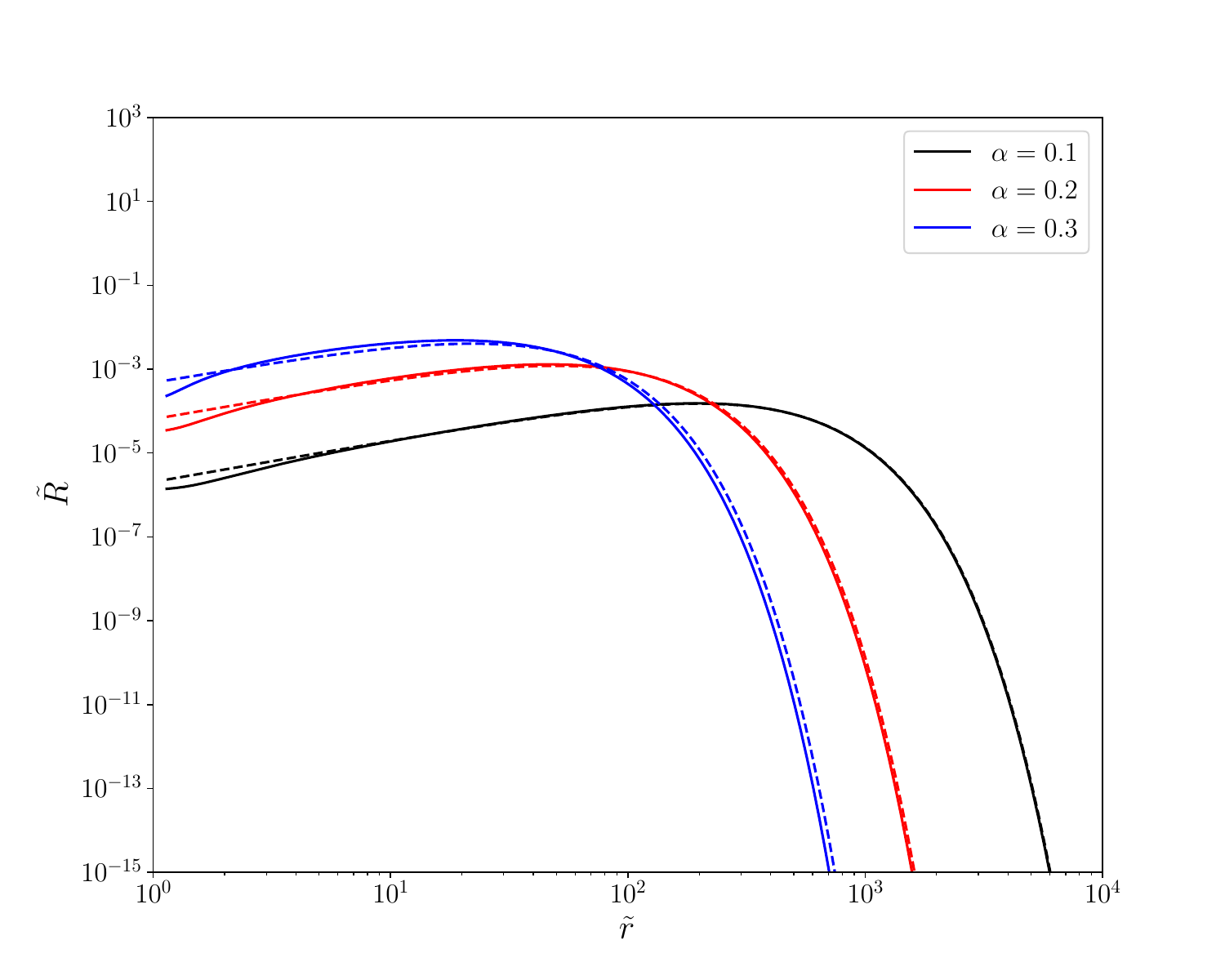}
    \caption{Comparison between the relativistic wavefunctions (solid lines) with the non-relativistic (dashed lines) for three different values of $\alpha$. We have set $\tilde{a} = 0.99$.\label{fig:relativistic}}
\end{figure}

Since our findings indicate that the signal is maximized for rather large values of $\alpha$, it is important to check if the non-relativistic approximation that we have employed throughout this paper holds and identify where it fails. Let us start with the comparison between the relativistic and non-relativistic wavefunctions for different values of $\alpha$. A deviation from the non-relativistic approximation would result in different values for the amplitude $\eta$ and hence different values of $z$. In order to check this effect, we used the code available in \cite{hoof2024gettingblackholesuperradiance,hoof_git} to calculate the energy of the $211$ level and combined this with the analysis outlined in \cite{witte2025steppingsuperradianceconstraintsaxions,Dolan:2007mj} to determine the normalized relativistic wavefunction $\tilde{R} = M^{3/2} R$ as a function of the distance $\tilde{r} = r/M$ from the black hole horizon. In \cref{fig:relativistic}, we show the relativistic wavefunctions for three different values of $\alpha$ and compare to their non-relativstic counterparts. As it is expected, the largest deviations are observed near the black hole horizon for all three cases, while the results converge at the $\tilde{r} \rightarrow \infty$ limit. In addition, as $\alpha$ increases, the relativistic wavefunction deviates more and more from the non-relativistic one, since the GA is much closer to the black hole horizon. However, these deviations appear to be small and will introduce an order one factor at the most.  

\begin{figure}[h!]
\centering
\includegraphics[width=0.7\linewidth]{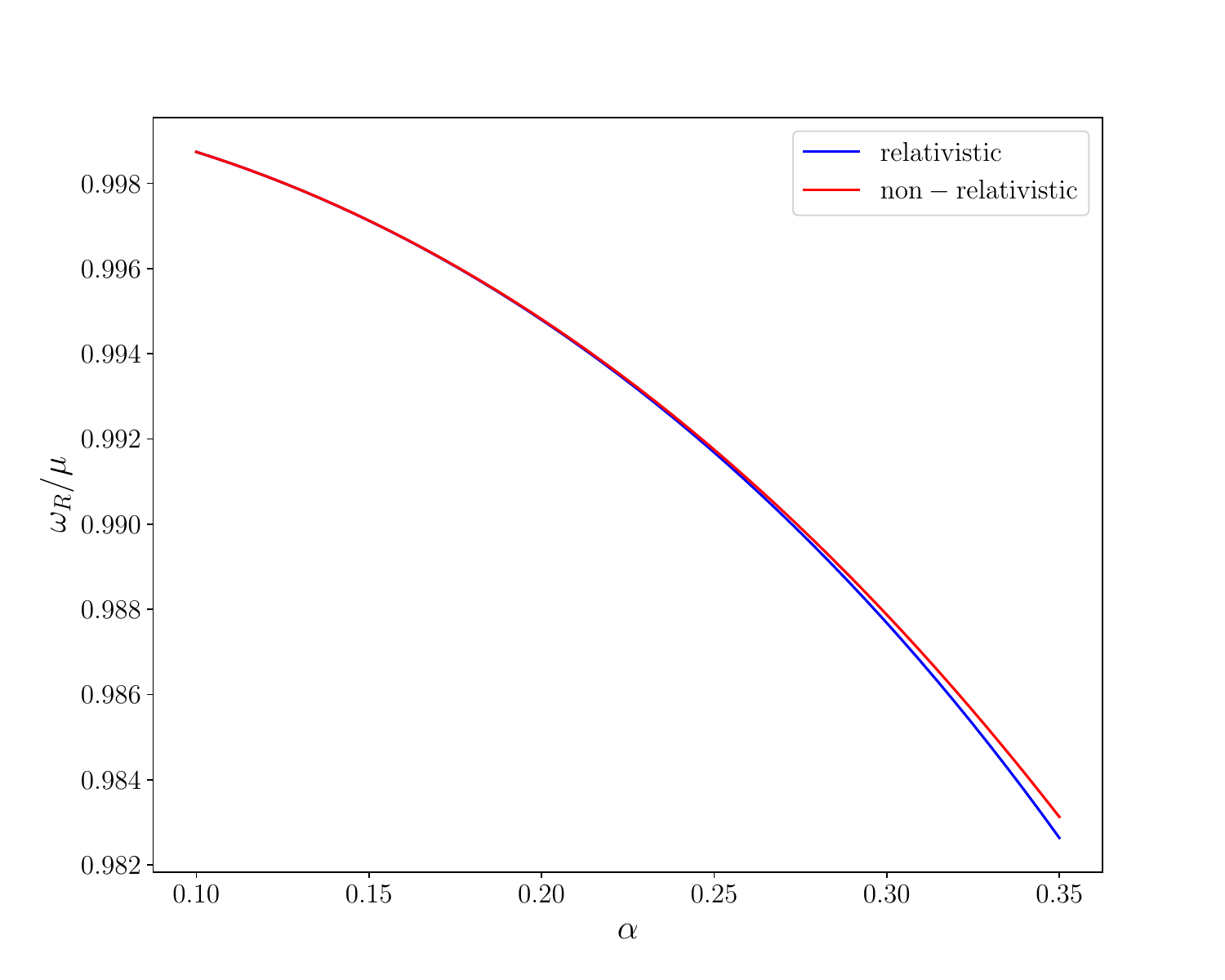}
    \caption{The real energy eigenvalue of the 211 level as a function of $\alpha$, using the relativistic and non-relativistic computations. The black hole's spin is set to $\tilde{a} = 0.99$}
    \label{fig:energy levels}
\end{figure}
\begin{figure}[h!]
\centering
\includegraphics[width=0.7\linewidth]{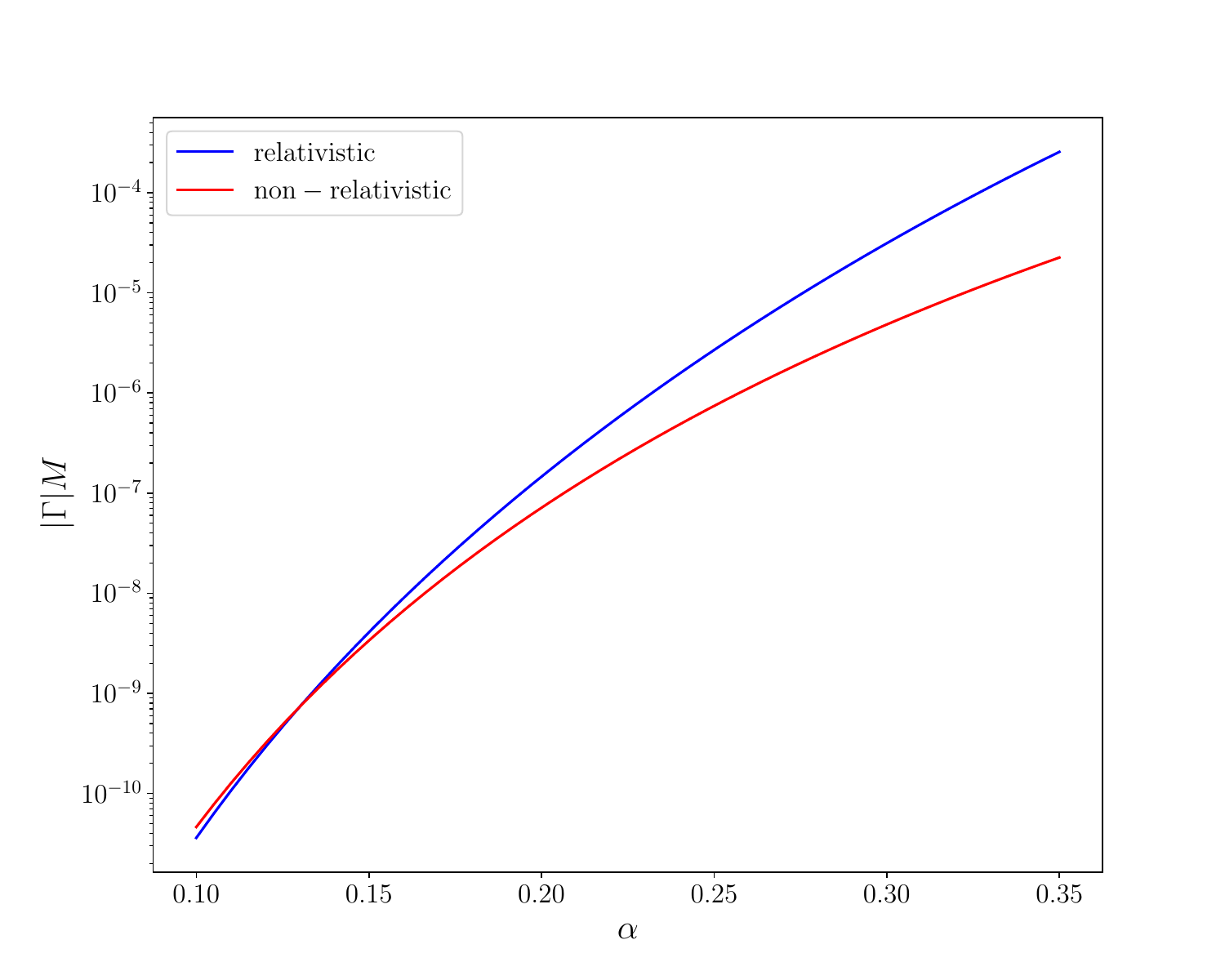}
    \caption{The decay rate of the second mode as a function of $\alpha$, comparing the relativistic and non-relativistic computations. The black hole's spin is set to the saturated spin $\tilde{a} = \frac{4\alpha}{1+4\alpha^{2}}$}
    \label{fig:decay rate plot}
\end{figure}
The same behavior shows up for the real part of the energy eigenvalue of the $211$ level, shown in \cref{fig:energy levels}. The relativistic solution starts to deviate from the non-relativistic one when $\alpha\gtrsim0.3$, but the overall deviation is negligible and should not affect our results. \par
The most important deviation of the relativistic results from the non-relativistic results is in the decay rate of the $|21\text{-}1\rangle$ state, shown in \cref{fig:decay rate plot}. As it is expected, the non-relativistic computation matches well with the relativistic result in the small $\alpha$ region, $\alpha\lesssim0.15$. However, the two results diverge at most by an order of magnitude at larger $\alpha\le0.35$. This has an important effect on the computation of the SNR, since it inversely proportional to $|\Gamma|$ in \cref{eqn:SNR mono}. Employing the non-relativistic approximation in this calculation would significantly overestimate the detectability prospects of the signal. 
\bibliographystyle{JHEP}
\bibliography{cit}

\providecommand{\href}[2]{#2}\begingroup\raggedright\begin{thebibliography}{10}

\bibitem{Brito_2020}
R.~Brito, V.~Cardoso and P.~Pani, \emph{{Superradiance}: {New Frontiers in Black Hole Physics}}, \href{https://doi.org/10.1007/978-3-319-19000-6}{\emph{Lect. Notes Phys.} {\bfseries 906} (2015) pp.1} [\href{https://arxiv.org/abs/1501.06570}{{\ttfamily 1501.06570}}].

\bibitem{Marsh_2016}
D.J.E.~Marsh, \emph{{Axion Cosmology}}, \href{https://doi.org/10.1016/j.physrep.2016.06.005}{\emph{Phys. Rept.} {\bfseries 643} (2016) 1} [\href{https://arxiv.org/abs/1510.07633}{{\ttfamily 1510.07633}}].

\bibitem{dark_matter}
M.~Cirelli, A.~Strumia and J.~Zupan, \emph{{Dark Matter}},  \href{https://arxiv.org/abs/2406.01705}{{\ttfamily 2406.01705}}.

\bibitem{Detweiler:1980uk}
S.L.~Detweiler, \emph{{KLEIN-GORDON EQUATION AND ROTATING BLACK HOLES}}, \href{https://doi.org/10.1103/PhysRevD.22.2323}{\emph{Phys. Rev. D} {\bfseries 22} (1980) 2323}.

\bibitem{spectra}
D.~Baumann, H.S.~Chia, J.~Stout and L.~ter Haar, \emph{{The Spectra of Gravitational Atoms}}, \href{https://doi.org/10.1088/1475-7516/2019/12/006}{\emph{JCAP} {\bfseries 12} (2019) 006} [\href{https://arxiv.org/abs/1908.10370}{{\ttfamily 1908.10370}}].

\bibitem{Arvanitaki_2011}
A.~Arvanitaki and S.~Dubovsky, \emph{{Exploring the String Axiverse with Precision Black Hole Physics}}, \href{https://doi.org/10.1103/PhysRevD.83.044026}{\emph{Phys. Rev. D} {\bfseries 83} (2011) 044026} [\href{https://arxiv.org/abs/1004.3558}{{\ttfamily 1004.3558}}].

\bibitem{Arvanitaki_2015}
A.~Arvanitaki, M.~Baryakhtar and X.~Huang, \emph{{Discovering the QCD Axion with Black Holes and Gravitational Waves}}, \href{https://doi.org/10.1103/PhysRevD.91.084011}{\emph{Phys. Rev. D} {\bfseries 91} (2015) 084011} [\href{https://arxiv.org/abs/1411.2263}{{\ttfamily 1411.2263}}].

\bibitem{_nal_2021}
C.~\"Unal, F.~Pacucci and A.~Loeb, \emph{{Properties of ultralight bosons from heavy quasar spins via superradiance}}, \href{https://doi.org/10.1088/1475-7516/2021/05/007}{\emph{JCAP} {\bfseries 05} (2021) 007} [\href{https://arxiv.org/abs/2012.12790}{{\ttfamily 2012.12790}}].

\bibitem{witte2025steppingsuperradianceconstraintsaxions}
S.J.~Witte and A.~Mummery, \emph{{Stepping Up Superradiance Constraints on Axions}},  \href{https://arxiv.org/abs/2412.03655}{{\ttfamily 2412.03655}}.

\bibitem{hoof2024gettingblackholesuperradiance}
S.~Hoof, D.J.E.~Marsh, J.~Sisk-Reyn\'es, J.H.~Matthews and C.~Reynolds, \emph{{Getting More Out of Black Hole Superradiance: a Statistically Rigorous Approach to Ultralight Boson Constraints}},  \href{https://arxiv.org/abs/2406.10337}{{\ttfamily 2406.10337}}.

\bibitem{mehta2021superradianceexclusionslandscapetype}
V.M.~Mehta, M.~Demirtas, C.~Long, D.J.E.~Marsh, L.~Mcallister and M.J.~Stott, \emph{{Superradiance Exclusions in the Landscape of Type IIB String Theory}},  \href{https://arxiv.org/abs/2011.08693}{{\ttfamily 2011.08693}}.

\bibitem{self-interactions}
M.~Baryakhtar, M.~Galanis, R.~Lasenby and O.~Simon, \emph{{Black hole superradiance of self-interacting scalar fields}}, \href{https://doi.org/10.1103/PhysRevD.103.095019}{\emph{Phys. Rev. D} {\bfseries 103} (2021) 095019} [\href{https://arxiv.org/abs/2011.11646}{{\ttfamily 2011.11646}}].

\bibitem{superradiance_string_theory}
V.M.~Mehta, M.~Demirtas, C.~Long, D.J.E.~Marsh, L.~McAllister and M.J.~Stott, \emph{{Superradiance in string theory}}, \href{https://doi.org/10.1088/1475-7516/2021/07/033}{\emph{JCAP} {\bfseries 07} (2021) 033} [\href{https://arxiv.org/abs/2103.06812}{{\ttfamily 2103.06812}}].

\bibitem{Gravitaitonal_wave_searches}
R.~Brito, S.~Ghosh, E.~Barausse, E.~Berti, V.~Cardoso, I.~Dvorkin et~al., \emph{{Gravitational wave searches for ultralight bosons with LIGO and LISA}}, \href{https://doi.org/10.1103/PhysRevD.96.064050}{\emph{Phys. Rev. D} {\bfseries 96} (2017) 064050} [\href{https://arxiv.org/abs/1706.06311}{{\ttfamily 1706.06311}}].

\bibitem{Yang_2023}
J.~Yang and F.P.~Huang, \emph{Gravitational waves from axions annihilation through quantum field theory}, \href{https://doi.org/10.1103/physrevd.108.103002}{\emph{Physical Review D} {\bfseries 108} (2023) }.

\bibitem{LIGOScientific:2021rnv}
{\scshape LIGO Scientific, Virgo, KAGRA} collaboration, \emph{{All-sky search for gravitational wave emission from scalar boson clouds around spinning black holes in LIGO O3 data}}, \href{https://doi.org/10.1103/PhysRevD.105.102001}{\emph{Phys. Rev. D} {\bfseries 105} (2022) 102001} [\href{https://arxiv.org/abs/2111.15507}{{\ttfamily 2111.15507}}].

\bibitem{deci_Hz}
H.~Omiya, T.~Takahashi, T.~Tanaka and H.~Yoshino, \emph{{Deci-Hz gravitational waves from the self-interacting axion cloud around a rotating stellar mass black hole}}, \href{https://doi.org/10.1103/PhysRevD.110.044002}{\emph{Phys. Rev. D} {\bfseries 110} (2024) 044002} [\href{https://arxiv.org/abs/2404.16265}{{\ttfamily 2404.16265}}].

\bibitem{Collaviti_2024}
S.~Collaviti, L.~Sun, M.~Galanis and M.~Baryakhtar, \emph{{Observational prospects of self-interacting scalar superradiance with next-generation gravitational-wave detectors}}, \href{https://doi.org/10.1088/1361-6382/ad96ff}{\emph{Class. Quant. Grav.} {\bfseries 42} (2025) 025006} [\href{https://arxiv.org/abs/2407.04304}{{\ttfamily 2407.04304}}].

\bibitem{DellaMonica:2025zby}
R.~Della~Monica and R.~Brito, \emph{{Detectability of gravitational atoms in black hole binaries with the Einstein Telescope}},  \href{https://arxiv.org/abs/2503.23419}{{\ttfamily 2503.23419}}.

\bibitem{Baumann_2019}
D.~Baumann, H.S.~Chia and R.A.~Porto, \emph{{Probing Ultralight Bosons with Binary Black Holes}}, \href{https://doi.org/10.1103/PhysRevD.99.044001}{\emph{Phys. Rev. D} {\bfseries 99} (2019) 044001} [\href{https://arxiv.org/abs/1804.03208}{{\ttfamily 1804.03208}}].

\bibitem{legacy}
G.M.~Tomaselli, T.F.M.~Spieksma and G.~Bertone, \emph{{Legacy of Boson Clouds on Black Hole Binaries}}, \href{https://doi.org/10.1103/PhysRevLett.133.121402}{\emph{Phys. Rev. Lett.} {\bfseries 133} (2024) 121402} [\href{https://arxiv.org/abs/2407.12908}{{\ttfamily 2407.12908}}].

\bibitem{Baumann_2020}
D.~Baumann, H.S.~Chia, R.A.~Porto and J.~Stout, \emph{{Gravitational Collider Physics}}, \href{https://doi.org/10.1103/PhysRevD.101.083019}{\emph{Phys. Rev. D} {\bfseries 101} (2020) 083019} [\href{https://arxiv.org/abs/1912.04932}{{\ttfamily 1912.04932}}].

\bibitem{resonant_history}
G.M.~Tomaselli, T.F.M.~Spieksma and G.~Bertone, \emph{{Resonant history of gravitational atoms in black hole binaries}}, \href{https://doi.org/10.1103/PhysRevD.110.064048}{\emph{Phys. Rev. D} {\bfseries 110} (2024) 064048} [\href{https://arxiv.org/abs/2403.03147}{{\ttfamily 2403.03147}}].

\bibitem{axion_cloud_backreaction}
T.~Takahashi, H.~Omiya and T.~Tanaka, \emph{{Axion cloud evaporation during inspiral of black hole binaries: The effects of backreaction and radiation}}, \href{https://doi.org/10.1093/ptep/ptac044}{\emph{PTEP} {\bfseries 2022} (2022) 043E01} [\href{https://arxiv.org/abs/2112.05774}{{\ttfamily 2112.05774}}].

\bibitem{Ionization}
D.~Baumann, G.~Bertone, J.~Stout and G.M.~Tomaselli, \emph{{Ionization of gravitational atoms}}, \href{https://doi.org/10.1103/PhysRevD.105.115036}{\emph{Phys. Rev. D} {\bfseries 105} (2022) 115036} [\href{https://arxiv.org/abs/2112.14777}{{\ttfamily 2112.14777}}].

\bibitem{sharp_signals}
D.~Baumann, G.~Bertone, J.~Stout and G.M.~Tomaselli, \emph{{Sharp Signals of Boson Clouds in Black Hole Binary Inspirals}}, \href{https://doi.org/10.1103/PhysRevLett.128.221102}{\emph{Phys. Rev. Lett.} {\bfseries 128} (2022) 221102} [\href{https://arxiv.org/abs/2206.01212}{{\ttfamily 2206.01212}}].

\bibitem{self_interaction_binary}
T.~Takahashi, H.~Omiya and T.~Tanaka, \emph{{Self-interacting axion clouds around rotating black holes in binary systems}}, \href{https://doi.org/10.1103/PhysRevD.110.104038}{\emph{Phys. Rev. D} {\bfseries 110} (2024) 104038} [\href{https://arxiv.org/abs/2408.08349}{{\ttfamily 2408.08349}}].

\bibitem{Guo:2024iye}
A.~Guo, J.~Zhang and H.~Yang, \emph{{Superradiant clouds may be relevant for close compact object binaries}}, \href{https://doi.org/10.1103/PhysRevD.110.023022}{\emph{Phys. Rev. D} {\bfseries 110} (2024) 023022} [\href{https://arxiv.org/abs/2401.15003}{{\ttfamily 2401.15003}}].

\bibitem{extreme_mass_ratio}
T.~Takahashi, H.~Omiya and T.~Tanaka, \emph{{Evolution of binary systems accompanying axion clouds in extreme mass ratio inspirals}}, \href{https://doi.org/10.1103/PhysRevD.107.103020}{\emph{Phys. Rev. D} {\bfseries 107} (2023) 103020} [\href{https://arxiv.org/abs/2301.13213}{{\ttfamily 2301.13213}}].

\bibitem{Zhang:2018kib}
J.~Zhang and H.~Yang, \emph{{Gravitational floating orbits around hairy black holes}}, \href{https://doi.org/10.1103/PhysRevD.99.064018}{\emph{Phys. Rev. D} {\bfseries 99} (2019) 064018} [\href{https://arxiv.org/abs/1808.02905}{{\ttfamily 1808.02905}}].

\bibitem{Bo_kovi__2024}
M.~Bo\v{s}kovi\'c, M.~Koschnitzke and R.A.~Porto, \emph{{Signatures of Ultralight Bosons in the Orbital Eccentricity of Binary Black Holes}}, \href{https://doi.org/10.1103/PhysRevLett.133.121401}{\emph{Phys. Rev. Lett.} {\bfseries 133} (2024) 121401} [\href{https://arxiv.org/abs/2403.02415}{{\ttfamily 2403.02415}}].

\bibitem{Tong_2022}
X.~Tong, Y.~Wang and H.-Y.~Zhu, \emph{{Termination of superradiance from a binary companion}}, \href{https://doi.org/10.1103/PhysRevD.106.043002}{\emph{Phys. Rev. D} {\bfseries 106} (2022) 043002} [\href{https://arxiv.org/abs/2205.10527}{{\ttfamily 2205.10527}}].

\bibitem{Carroll:2004st}
S.M.~Carroll, \emph{{Spacetime and Geometry}: {An Introduction to General Relativity}}, Cambridge University Press (7, 2019), \href{https://doi.org/10.1017/9781108770385}{10.1017/9781108770385}.

\bibitem{hoof_git}
S.~Hoof. \url{https://github.com/sebhoof/bhsr}.

\bibitem{axion_annihilation}
J.~Yang and F.P.~Huang, \emph{{Gravitational waves from axions annihilation through quantum field theory}}, \href{https://doi.org/10.1103/PhysRevD.108.103002}{\emph{Phys. Rev. D} {\bfseries 108} (2023) 103002} [\href{https://arxiv.org/abs/2306.12375}{{\ttfamily 2306.12375}}].

\bibitem{Landau}
L.D.~Landau, \emph{{A theory of energy transfer. 2.}}, \href{https://doi.org/10.1016/B978-0-08-010586-4.50014-6}{\emph{Phys. Z. Sowjetunion} {\bfseries 2} (1932) }.

\bibitem{Zener}
C.~Zener, \emph{{Nonadiabatic crossing of energy levels}}, \href{https://doi.org/10.1098/rspa.1932.0165}{\emph{Proc. Roy. Soc. Lond. A} {\bfseries 137} (1932) 696}.

\bibitem{hyperfine_trans_are_favoured}
X.~Tong, Y.~Wang and H.-Y.~Zhu, \emph{{Gravitational Collider Physics via Pulsar\textendash{}Black Hole Binaries II: Fine and Hyperfine Structures Are Favored}}, \href{https://doi.org/10.3847/1538-4357/ac36db}{\emph{Astrophys. J.} {\bfseries 924} (2022) 99} [\href{https://arxiv.org/abs/2106.13484}{{\ttfamily 2106.13484}}].

\bibitem{D}
N.M.~Temme, ``{NIST Digital Library of Mathematical Functions}.''

\bibitem{Siemonsen:2022yyf}
N.~Siemonsen, T.~May and W.E.~East, \emph{{Modeling the black hole superradiance gravitational waveform}}, \href{https://doi.org/10.1103/PhysRevD.107.104003}{\emph{Phys. Rev. D} {\bfseries 107} (2023) 104003} [\href{https://arxiv.org/abs/2211.03845}{{\ttfamily 2211.03845}}].

\bibitem{decayrate1}
V.M.~Akulin and W.P.~Schleich, \emph{Landau-zener transition to a decaying level}, \href{https://doi.org/10.1103/PhysRevA.46.4110}{\emph{Phys. Rev. A} {\bfseries 46} (1992) 4110}.

\bibitem{decayrate2}
N.V.~Vitanov and S.~Stenholm, \emph{Pulsed excitation of a transition to a decaying level}, \href{https://doi.org/10.1103/PhysRevA.55.2982}{\emph{Phys. Rev. A} {\bfseries 55} (1997) 2982}.

\bibitem{Shankar:102017}
R.~Shankar, \emph{{Principles of quantum mechanics}}, Plenum, New York, NY (1980).

\bibitem{Maggiore:2007ulw}
M.~Maggiore, \emph{{Gravitational Waves. Vol. 1: Theory and Experiments}}, Oxford University Press (2007), \href{https://doi.org/10.1093/acprof:oso/9780198570745.001.0001}{10.1093/acprof:oso/9780198570745.001.0001}.

\bibitem{May:2024npn}
T.~May, W.E.~East and N.~Siemonsen, \emph{{Self-Gravity Effects of Ultralight Boson Clouds Formed by Black Hole Superradiance}},  \href{https://arxiv.org/abs/2410.21442}{{\ttfamily 2410.21442}}.

\bibitem{Droz:1999qx}
S.~Droz, D.J.~Knapp, E.~Poisson and B.J.~Owen, \emph{{Gravitational waves from inspiraling compact binaries: Validity of the stationary phase approximation to the Fourier transform}}, \href{https://doi.org/10.1103/PhysRevD.59.124016}{\emph{Phys. Rev. D} {\bfseries 59} (1999) 124016} [\href{https://arxiv.org/abs/gr-qc/9901076}{{\ttfamily gr-qc/9901076}}].

\bibitem{Yoshino:2013ofa}
H.~Yoshino and H.~Kodama, \emph{{Gravitational radiation from an axion cloud around a black hole: Superradiant phase}}, \href{https://doi.org/10.1093/ptep/ptu029}{\emph{PTEP} {\bfseries 2014} (2014) 043E02} [\href{https://arxiv.org/abs/1312.2326}{{\ttfamily 1312.2326}}].

\bibitem{Tomaselli:2024faa}
G.M.~Tomaselli, \emph{{Gravitational atoms and black hole binaries}}, Ph.D. thesis, Amsterdam U., 2024.
\newblock \href{https://arxiv.org/abs/2412.12526}{{\ttfamily 2412.12526}}.

\bibitem{LISA:2024hlh}
{\scshape LISA} collaboration, \emph{{LISA Definition Study Report}},  \href{https://arxiv.org/abs/2402.07571}{{\ttfamily 2402.07571}}.

\bibitem{DECIGO}
K.~Yagi and N.~Seto, \emph{{Detector configuration of DECIGO/BBO and identification of cosmological neutron-star binaries}}, \href{https://doi.org/10.1103/PhysRevD.83.044011}{\emph{Phys. Rev. D} {\bfseries 83} (2011) 044011} [\href{https://arxiv.org/abs/1101.3940}{{\ttfamily 1101.3940}}].

\bibitem{MacLeod:2015bpa}
M.~MacLeod, M.~Trenti and E.~Ramirez-Ruiz, \emph{{The Close Stellar Companions to Intermediate Mass Black Holes}}, \href{https://doi.org/10.3847/0004-637X/819/1/70}{\emph{Astrophys. J.} {\bfseries 819} (2016) 70} [\href{https://arxiv.org/abs/1508.07000}{{\ttfamily 1508.07000}}].

\bibitem{askar2024intermediatemassblackholesstar}
A.~Askar, V.F.~Baldassare and M.~Mezcua, \emph{{Intermediate-Mass Black Holes in Star Clusters and Dwarf Galaxies}},  11, 2023 [\href{https://arxiv.org/abs/2311.12118}{{\ttfamily 2311.12118}}].

\bibitem{Berti:2008af}
E.~Berti and M.~Volonteri, \emph{{Cosmological black hole spin evolution by mergers and accretion}}, \href{https://doi.org/10.1086/590379}{\emph{Astrophys. J.} {\bfseries 684} (2008) 822} [\href{https://arxiv.org/abs/0802.0025}{{\ttfamily 0802.0025}}].

\bibitem{Borchers:2025sid}
A.~Borchers, C.S.~Ye and M.~Fishbach, \emph{{Gravitational-wave kicks impact spins of black holes from hierarchical mergers}},  \href{https://arxiv.org/abs/2503.21278}{{\ttfamily 2503.21278}}.

\bibitem{LIGOScientific:2025rsn}
{\scshape LIGO Scientific, VIRGO, KAGRA} collaboration, \emph{{GW231123: a Binary Black Hole Merger with Total Mass 190-265 $M_{\odot}$}},  \href{https://arxiv.org/abs/2507.08219}{{\ttfamily 2507.08219}}.

\bibitem{Arca-Sedda:2020lso}
M.~Arca-Sedda, P.~Amaro-Seoane and X.~Chen, \emph{{Merging stellar and intermediate-mass black holes in dense clusters: implications for LIGO, LISA, and the next generation of gravitational wave detectors}}, \href{https://doi.org/10.1051/0004-6361/202037785}{\emph{Astron. Astrophys.} {\bfseries 652} (2021) A54} [\href{https://arxiv.org/abs/2007.13746}{{\ttfamily 2007.13746}}].

\bibitem{10.1093/mnras/stw1779}
A.~Sollima, F.R.~Ferraro, L.~Lovisi, F.~Contenta, E.~Vesperini, L.~Origlia et~al., \emph{Searching in the dark: the dark mass content of the milky way globular clusters ngc288 and ngc6218}, \href{https://doi.org/10.1093/mnras/stw1779}{\emph{Monthly Notices of the Royal Astronomical Society} {\bfseries 462} (2016) 1937}.

\bibitem{Kiziltan:2017ijz}
B.~K{\i}z{\i}ltan, H.~Baumgardt and A.~Loeb, \emph{{An intermediate-mass black hole in the centre of the globular cluster 47 Tucanae}}, \href{https://doi.org/10.1038/nature21361}{\emph{Nature} {\bfseries 542} (2017) 203} [\href{https://arxiv.org/abs/1702.02149}{{\ttfamily 1702.02149}}].

\bibitem{Abbate_2019}
F.~Abbate, A.~Possenti, M.~Colpi and M.~Spera, \emph{Evidence of nonluminous matter in the center of m62}, \href{https://doi.org/10.3847/2041-8213/ab46c3}{\emph{The Astrophysical Journal Letters} {\bfseries 884} (2019) L9}.

\bibitem{2019MNRAS.487.2685E}
D.~{Erkal}, V.~{Belokurov}, C.F.P.~{Laporte}, S.E.~{Koposov}, T.S.~{Li}, C.J.~{Grillmair} et~al., \emph{{The total mass of the Large Magellanic Cloud from its perturbation on the Orphan stream}}, \href{https://doi.org/10.1093/mnras/stz1371}{\emph{\mnras} {\bfseries 487} (2019) 2685} [\href{https://arxiv.org/abs/1812.08192}{{\ttfamily 1812.08192}}].

\bibitem{2017ApJ...846...14B}
H.~{Boyce}, N.~{L{\"u}tzgendorf}, R.P.~{van der Marel}, H.~{Baumgardt}, M.~{Kissler-Patig}, N.~{Neumayer} et~al., \emph{{An Upper Limit on the Mass of a Central Black Hole in the Large Magellanic Cloud from the Stellar Rotation Field}}, \href{https://doi.org/10.3847/1538-4357/aa830c}{\emph{\apj} {\bfseries 846} (2017) 14} [\href{https://arxiv.org/abs/1612.00045}{{\ttfamily 1612.00045}}].

\bibitem{2007MNRAS.376L..29G}
A.~{Gualandris} and S.~{Portegies Zwart}, \emph{{A hypervelocity star from the Large Magellanic Cloud}}, \href{https://doi.org/10.1111/j.1745-3933.2007.00280.x}{\emph{\mnras} {\bfseries 376} (2007) L29} [\href{https://arxiv.org/abs/astro-ph/0612673}{{\ttfamily astro-ph/0612673}}].

\bibitem{Peng:2025zca}
S.-T.~Peng and J.~Zhang, \emph{{Gravitational Waves from Superradiant Cloud Level Transition}},  \href{https://arxiv.org/abs/2504.00728}{{\ttfamily 2504.00728}}.

\bibitem{Cheung:2025grp}
M.H.-Y.~Cheung, D.~Wadekar, A.K.~Mehta, T.~Islam, J.~Roulet, E.~Berti et~al., \emph{{Searching for intermediate mass ratio binary black hole mergers in the third observing run of LIGO-Virgo-KAGRA}},  \href{https://arxiv.org/abs/2507.01083}{{\ttfamily 2507.01083}}.

\bibitem{Berti:2019wnn}
E.~Berti, R.~Brito, C.F.B.~Macedo, G.~Raposo and J.L.~Rosa, \emph{{Ultralight boson cloud depletion in binary systems}}, \href{https://doi.org/10.1103/PhysRevD.99.104039}{\emph{Phys. Rev. D} {\bfseries 99} (2019) 104039} [\href{https://arxiv.org/abs/1904.03131}{{\ttfamily 1904.03131}}].

\bibitem{Dandoy:2022prp}
V.~Dandoy, T.~Schwetz and E.~Todarello, \emph{{A self-consistent wave description of axion miniclusters and their survival in the galaxy}}, \href{https://doi.org/10.1088/1475-7516/2022/09/081}{\emph{JCAP} {\bfseries 09} (2022) 081} [\href{https://arxiv.org/abs/2206.04619}{{\ttfamily 2206.04619}}].

\bibitem{Li:2025ffh}
D.~Li, C.~Weller, P.~Bourg, M.~LaHaye, N.~Yunes and H.~Yang, \emph{{Extreme mass-ratio inspiral within an ultralight scalar cloud I. Scalar radiation}},  \href{https://arxiv.org/abs/2507.02045}{{\ttfamily 2507.02045}}.

\bibitem{Dolan:2007mj}
S.R.~Dolan, \emph{{Instability of the massive Klein-Gordon field on the Kerr spacetime}}, \href{https://doi.org/10.1103/PhysRevD.76.084001}{\emph{Phys. Rev. D} {\bfseries 76} (2007) 084001} [\href{https://arxiv.org/abs/0705.2880}{{\ttfamily 0705.2880}}].

\end{thebibliography}\endgroup

\end{document}